\documentclass[10pt,twocolumn,letterpaper]{article}
\usepackage[pagenumbers]{cvpr} 

\usepackage{algorithm}
\usepackage{algpseudocode}
\usepackage{amsfonts}
\usepackage{multirow}
\usepackage{float}
\usepackage{amssymb} 
\usepackage{graphicx} 
\usepackage{url}
\graphicspath{{figs/}} 
\usepackage{pifont}

\usepackage{colortbl}
\definecolor{grey1}{gray}{0.85}
\definecolor{greyG}{rgb}{0.94, 0.88, 0.88}  
\definecolor{greyR}{rgb}{0.88, 0.94, 0.88}
\definecolor{greyB}{rgb}{0.96, 0.96, 0.85}
\definecolor{cvprblue}{rgb}{0.21,0.49,0.74}
\usepackage[pagebackref,breaklinks,colorlinks,allcolors=cvprblue]{hyperref}
\usepackage{subcaption}
\usepackage{graphicx}
\usepackage{float}
\usepackage{cuted}
\usepackage{capt-of} 
\usepackage[switch]{lineno}

\title{Performance evaluation of scheduling tasks in many-core systems utilizing  processes and threads}
\author{Mejgan Dedaj\\
	MicroComputer Systems Laboratory\\
	Dept. of Mathematics, University of Ioannina\\
	{\tt\small dedajmegi@gmail.com}
	\and
	Argyro Gailla\\
	MicroComputer Systems Laboratory\\
	Dept. of Mathematics, University of Ioannina\\
	{\tt\small silvigail@hotmail.com}
	\and
	Theofanis Ioannou\\
	MicroComputer Systems Laboratory\\
	Dept. of Mathematics, University of Ioannina\\
	{\tt\small th.ioannou01@gmail.com}
	\and
	Stamatia Kastrinaki\\
	MicroComputer Systems Laboratory\\
	Dept. of Mathematics, University of Ioannina\\
	{\tt\small matinakst@gmail.com}
	\and
	Hermione Kimpouropoulou\\
	MicroComputer Systems Laboratory\\
	Dept. of Mathematics, University of Ioannina\\
	{\tt\small hermionekimp@gmail.com  }
	\and
	Dimitrios Kontodimos\\
	MicroComputer Systems Laboratory\\
	Dept. of Mathematics, University of Ioannina\\
	{\tt\small Jimkon03@gmail.com}
	\and
	Kleopatra Kontogianni\\
	MicroComputer Systems Laboratory\\
	Dept. of Mathematics, University of Ioannina\\
	{\tt\small kontkleo13@yahoo.gr}
	\and
	Sotirios Kontogiannis*\\
	MicroComputer Systems Laboratory\\
	Dept. of Mathematics, University of Ioannina\\
	{\tt\small skontog@uoi.gr}
	\and
	Michail Panagiotidis Kannas\\
	MicroComputer Systems Laboratory\\
	Dept. of Mathematics, University of Ioannina\\
	{\tt\small kannasmichael@gmail.com }
	\and
	Anastasia Papouda\\
	MicroComputer Systems Laboratory\\
	Dept. of Mathematics, University of Ioannina\\
	{\tt\small natasapapouda@gmail.com}
	\and
	Anna Maria Sidiropoulou\\
	MicroComputer Systems Laboratory\\
	Dept. of Mathematics, University of Ioannina\\
	{\tt\small annasidi360@gmail.com}
	\and
	George Tavridis\\
	MicroComputer Systems Laboratory\\
	Dept. of Mathematics, University of Ioannina\\
	{\tt\small georgetavridis7@gmail.com}
}
\begin{document}
\maketitle
\def\thefootnote{}\footnotetext{These authors contributed equally to this work.}
\def\thefootnote{*}\footnotetext{Correspondence to: skontog@uoi.gr}
\begin{abstract}
	This study assesses the scalability of process-based and thread-based schedulers for many-core shared-memory systems using a memory-intensive row-wise quick-sort workload on large three-dimensional tensors. The process-based evaluation considers bounded prolific, bounded collective, and three pipe-based producer-consumer schedulers: one-to-one, one-to-many, and many-to-many. These pipe schedulers dynamically stream task identifiers to worker processes, exchanging increased inter-process communication overhead for enhanced runtime load balancing and flexible chunk-based task dispatching. The thread-based evaluation examines static, dynamic, guided, chunk-based, chunk-stealing, adaptive chunk, and AIMD adaptive scheduling strategies. The AIMD scheduler employs an additive-increase multiplicative-decrease policy inspired by TCP congestion control, utilizing an exponentially weighted moving average (EWMA) of CPU utilization to regulate a contention window that limits the number of concurrently active chunks. The adaptive chunk scheduler further modifies chunk size based on observed per-thread execution speed. Experimental results on a 24-core x86-64 platform indicate that thread schedulers deliver the highest overall performance, with dynamic and guided scheduling yielding the most favorable practical outcomes. Among process schedulers, pipe-based designs demonstrate the strongest scalability, with one-to-one pipes excelling for smaller workloads and many-to-many pipes preferred for larger workloads. In summary, lightweight thread scheduling is optimal for shared-memory row sorting, while AIMD/adaptive scheduling and pipe-based process scheduling remain valuable for contention-aware execution, explicit inter-process coordination, and distributed-style heterogeneous workload management.
\end{abstract}    
\section{Introduction}
\label{sec:intro}

Process-based scheduling is a central perspective for heavyweight tasks executed in a Single Instruction Multiple Process (SIMP) manner. In prolific forking, the parent process is used if its PID is a first-degree descendant of the parent process, and each child process, before carrying out any assigned computation, increases the contention in an additive manner \cite{vasileiadou25}. In collective forking, the parent process stores the PIDs of all spawned child processes in a vector and sets itself as a subreaper or process group leader for all children \cite{vasileiadou25}. 

Since process forking uses Copy-on-Write (COW), the use of shared memory segments or shared memory maps is essential for task-data allocations in a SIMP manner \cite{bovet2005}. This shared segment is considered critical, and parallel read-writes to the same memory space should be avoided or serialized using mutex locks \cite{bovet2005}. According to \cite{vasileiadou25}, the bounded n-process collective forking is considered better than the prolific approach in terms of execution time for parallel tensor multiplications executed on x86\_64 multi-core CPUs, where n is the number of the system's available cores. 

On ARM quad-core systems, the bounded 100-processes prolific forking performs the best \cite{vasileiadou25}. Furthermore, in \cite{vasileiadou25}, experimentation shows that prolific forking performs better for low contention, while collective forking performs better for medium and high contention \cite{Sysel20}. Therefore, collective forking is generally better for heavier process workloads, while prolific forking can be better for lighter workloads, especially on ARM \cite{vasileiadou25}. Additionally, a collective scheduling with mutation is a genetic-algorithm (GA) scheduler in which an entire set of jobs is encoded as a collective task set, and the mutation operator randomly changes the assignment or order of selected tasks to explore better machine-load balance, lower completion time, or improved resource utilization \cite{Lawrynowicz11}.

Thread scheduling, in contrast, is more naturally associated with shared-memory execution models. In this work, rather than using the actual OpenMP runtime system, we implement several OpenMP-inspired scheduling policies—such as static, dynamic, and guided chunk allocation—strictly inside our shared-memory framework using POSIX threads and the Boost library \cite{OpenMP52Spec, posix2017}. In these OpenMP-inspired policies, work is commonly distributed among threads through scheduling policies such as static, dynamic, and guided, while the runtime schedule allows the actual policy to be selected at execution time through the environment or runtime system \cite{OpenMP52Spec, thoman12}. Dynamic scheduling improves load balance by assigning work to threads on demand, whereas guided scheduling starts with larger chunks and gradually reduces them to lower overhead while preserving balance \cite{OpenMP52Spec, thoman12, muller25}. 

Beyond these built-in mechanisms, adaptive thread schedulers extend these OpenMP-inspired policies by adjusting chunk sizes or task assignment decisions according to observed runtime behavior, such as thread speed, queue lengths, or processor utilization \cite{akturk2019}. In this sense, dynamic and runtime-inspired approaches provide flexibility within this framework, while adaptive schedulers aim to improve robustness and efficiency under irregular workloads, all realized natively via Boost threads \cite{agathos22, thoman12}. For irregular applications, adaptive schedulers try to reduce both load imbalance and scheduling overhead by dynamically changing chunk sizes and enabling work stealing / task stealing when a worker becomes idle \cite{blumofe1999}. Towards this direction, Booth and Lane \cite{booth22} proposed an adaptive self-scheduling method that self-manages chunk size based on a heuristic of workload variance and uses work stealing \cite{blumofe1999}. At the same time, Guo et al. introduced SLAW \cite{guo10}. This adaptive work-stealing scheduler selects a scheduling policy at runtime and improves locality by restricting stealing in a structured way \cite{guo10}. Together, these approaches show that adaptive chunk scheduling combined with stealing mechanisms can improve throughput and robustness compared with fixed scheduling policies, especially for irregular or imbalanced workloads \cite{Mohtasham15}.

\section{Materials and methods}
Focusing on the characteristics of processes and threads in multi-core systems, this paper evaluates the potential of different process and thread schedulers \cite{Sysel20, vasileiadou25}. The following sections present the experimental schedulers and performance evaluation metrics used in this study. 

\subsection{Process schedulers}
Process policies primarily adapt to inherent attributes that favor high-performance computations in SIMP \cite{gropp1999usingmpi, vasileiadou25}. That is a common data memory segment, a common data results segment, and a bounded pool of k processes , typically fewer than the number of CPU cores assigned to a specific workload \cite{vasileiadou25}. Nevertheless, compared with threads, processes are typically heavier-weight execution units because child processes are commonly created via fork(), which uses copy-on-write memory duplication and incurs separate address-space management costs \cite{bovet2005}. In contrast, threads execute within a shared address space and therefore usually exhibit lower creation and scheduling overhead \cite{Sysel20}.

Both the prolific and collective schedulers can be interpreted through a mitosis-like process-creation model, since child processes are created through repeated fork operations. In this analogy, each process fork represents process mitosis, in which a process duplicates itself and creates a new child process.

The difference between the two schedulers lies in their process collection policy. A prolific scheduler keeps only the direct child processes of the original main parent process. Therefore, for (n) direct fork operations issued by the main parent, the prolific scheduler manages (n) direct children and does not manage descendant-generated processes as workers.

A collective scheduler, in contrast, relies on the same mitosis-like forking mechanism, but it may keep the child processes produced during the complete forking evolution, regardless of their parent process. Since ($n$) fork levels may generate up to $(2^n)$ total processes, the collective scheduler may manage up to $(2^n - 1)$ child processes. Thus, both schedulers are based on process mitosis, but the prolific scheduler preserves only the main parent’s direct lineage, whereas the collective scheduler manages the wider fork-generated population.

However, gathering processes in the collective scheduler introduces additional complexity because child processes may originate from different parent processes and may therefore have different execution contexts, states, priorities, lifetimes, and resource requirements. Unlike the prolific scheduler, which tracks only the direct children of the main parent process, the collective scheduler must discover, identify, synchronize, and manage a larger and more heterogeneous population of fork-generated processes. This requires additional effort to maintain process identifiers, enforce process group IDs, use subreaper mechanisms, or store process identifiers in shared-memory lists or vectors. As a result, collective process management may increase coordination overhead and introduce synchronization delays, race conditions, orphan or zombie process management issues, and higher memory and CPU costs for termination or barrier synchronization.

The drawback of the prolific scheduler is that it manages only the direct children of the main parent process. This makes the scheduler simpler and easier to control, but it reduces the available degree of parallelism compared with a fully collective fork-generated process population and may lead to underutilization of potential workers.

The assignment of tasks among child processes should be considered part of the mapping policy. Workload decomposition corresponds to the partitioning of the original workload into distinct computational parts, represented here by worker processes. Once the workload has been divided into chunks, the scheduler applies a mapping strategy to assign each task or chunk to a specific child process and, when affinity is enabled, to a specific CPU core. In oversubscribed configurations, where the number of workers $(k)$ exceeds the number of available cores ($N$), worker-to-core assignment may follow a round-robin policy.

\subsubsection{Bounded prolific scheduler} 

The bounded prolific scheduler follows a parent-centric process creation strategy in which the main process directly spawns all worker processes using  \texttt{fork()}. Work is distributed cyclically (in a round-robin manner) among workers, assigning task $i$ to worker $i \bmod k$, where $k$ denotes the number of active workers. 

Because all workers are created directly by the parent, the process tree remains shallow and easy to manage. However, process creation pressure accumulates on the parent process, potentially increasing contention as the number of workers grows. The scheduler therefore favors workloads with relatively stable and balanced task distributions, where its low coordination overhead can outweigh the cost of centralized process creation. 

\subsubsection{Bounded collective scheduler} 
The bounded collective scheduler extends collective process creation by organizing 
worker generation under the fork mitosis operation. Instead of spawning workers linearly, 
each process recursively forks additional workers according to logarithmic expansion levels, 
reducing the depth of the spawning hierarchy. 
For $k$ workers, the scheduler requires approximately $log_2(k)$ spawning levels. This 
structure attempts to distribute process-creation overhead more evenly across workers and 
reduce parent-process contention for large worker counts.

\subsubsection{Pipe Process schedulers}

In this paper, an additional process scheduler for streaming workloads to processes via pipes using the producer-bounded consumers (RTP/C) paradigm is also evaluated \cite{jeffay93, dijkstra68, hoare1978, posix2017}. In these schedulers, workload 
identifiers are streamed dynamically between producer and worker processes through :
\begin{itemize}
\item The \textbf{one-to-one scheduler} uses a single communication pipe shared by all worker processes. The parent process writes tasks to the pipe, while workers read and execute the assigned tasks from the same communication channel.
 
\item The \textbf{one-to-many scheduler} uses multiple communication pipes, one for each worker process. The parent process distributes tasks directly to individual workers through dedicated communication channels, allowing independent task assignment and reducing contention among workers.
	
	\item The \textbf{many-to-many scheduler} uses separate communication channels for task assignment and completion notifications. Tasks are distributed from the parent process to workers through write pipes, while workers report completion through dedicated done pipes, enabling bidirectional communication and improved scheduling flexibility.

\end{itemize}
Compared with static process-assignment strategies, pipe-based schedulers introduce additional IPC (Inter-Process Communication) overhead, since task identifiers must be transferred through pipe read/write operations.

However, this extra communication cost may be compensated by improved runtime balancing, especially when the workload is irregular or when workers finish their assigned tasks at different times. 

Additionally, all three pipe-based schedulers are strongly affected by the selected chunk size. In this implementation, each task corresponds to one frame of the input tensor, and a chunk represents a group of consecutive frame identifiers streamed through the pipe as a single scheduling unit. Therefore, when $(chunk = 1)$, each worker receives only one frame at a time. This provides the finest scheduling granularity and usually improves load balancing, because idle workers can quickly request new work. However, it also increases the number of pipe transactions, system calls, and scheduling decisions, which may reduce efficiency when the per-frame computation is small or relatively uniform.

\begin{figure}[ht] 
	\centering
	\begin{subfigure}[b]{0.75\columnwidth} 
		\centering
		\includegraphics[width=\textwidth]{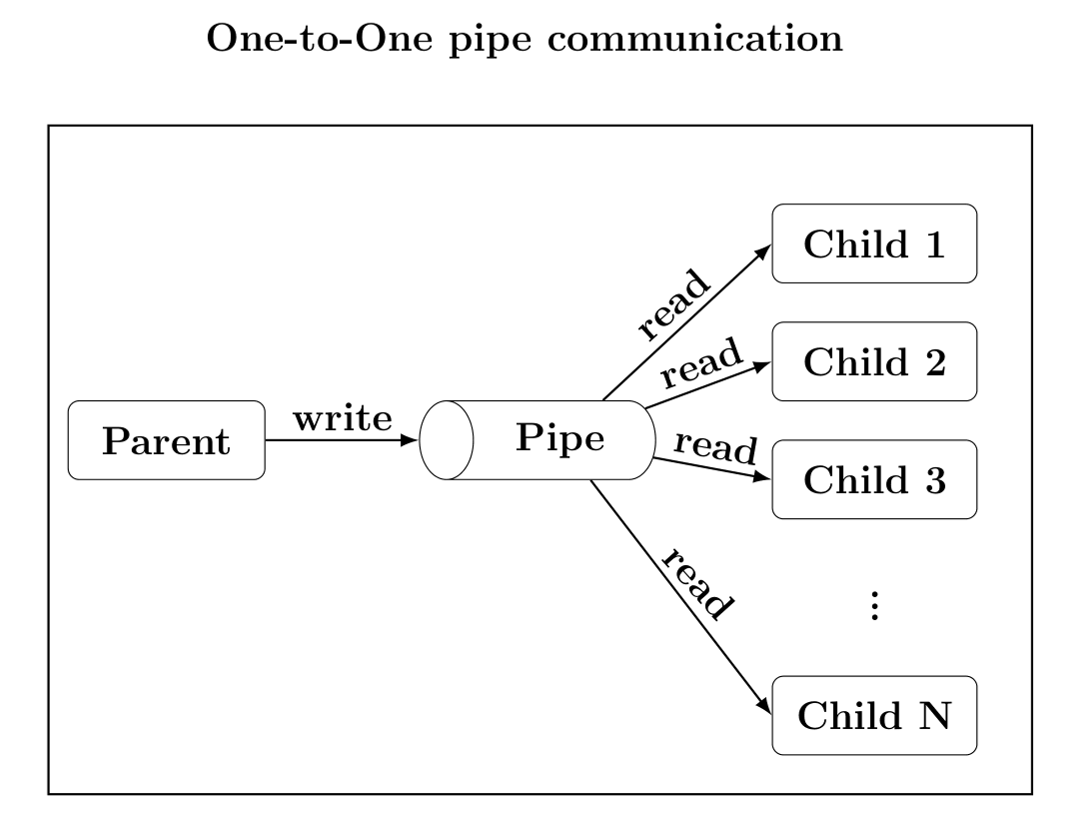}
		\caption{Pipe One-to-One}
		\label{subfig:one_to_one}
	\end{subfigure}
	\begin{subfigure}[b]{0.75\columnwidth}
		\centering
		\includegraphics[width=\textwidth]{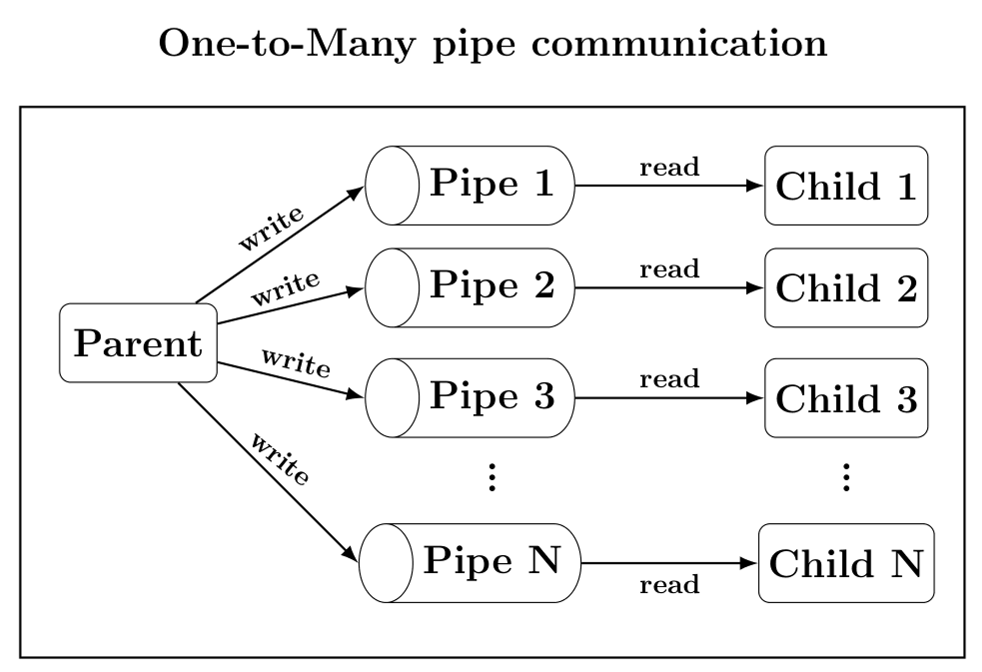}
		\caption{Pipe One-to-Many}
		\label{subfig:one_to_many}
	\end{subfigure}
	\begin{subfigure}[b]{0.75\columnwidth}
		\centering
		\includegraphics[width=\textwidth]{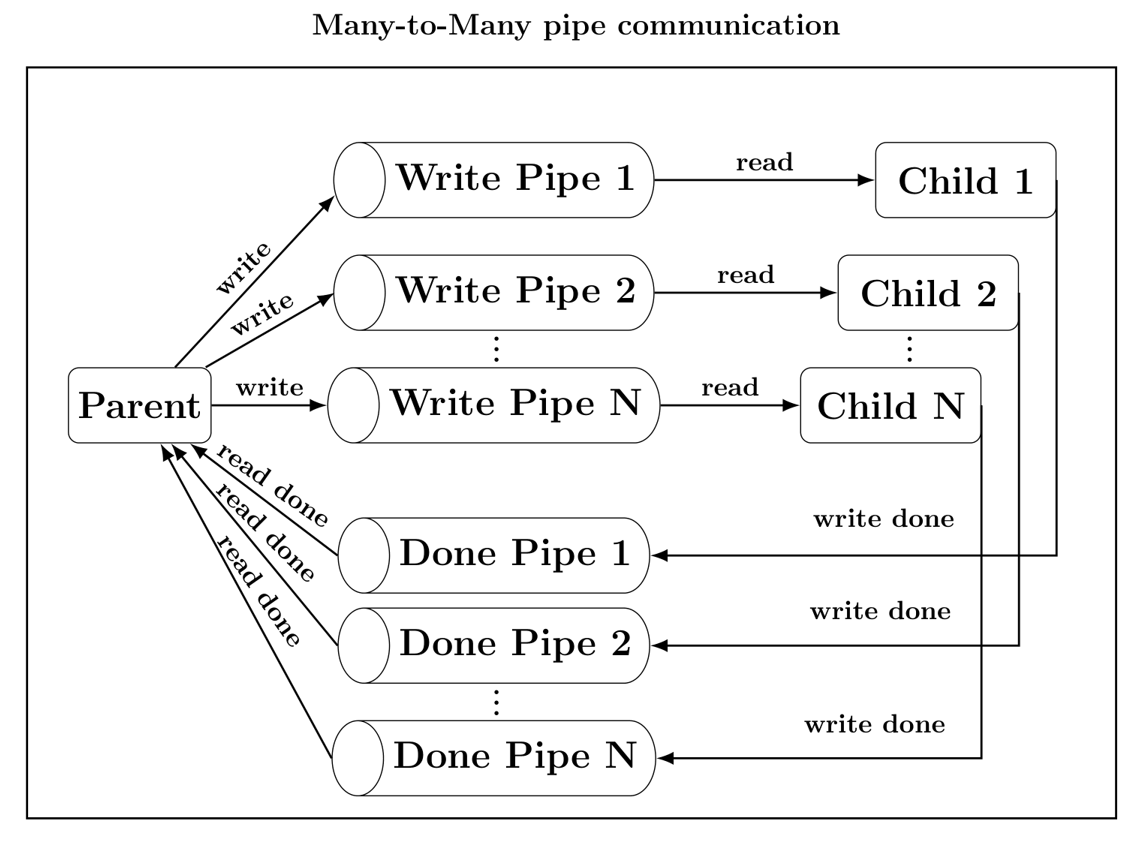}
		\caption{Pipe Many-to-Many}
		\label{subfig:many_to_many}
	\end{subfigure}
	\caption{Comparison of Inter-Process Communication (IPC) Pipe Configurations.}
	\label{fig:pipe_configurations}
\end{figure}

Larger chunk values reduce the frequency of communication through the pipes, because each worker receives a batch of frames per read operation rather than a single frame. This lowers IPC and scheduling overhead, improving efficiency when the workload is sufficiently regular. However, if the chunk size becomes too large, load imbalance may appear: some workers may receive heavier batches and continue executing while others become idle. Thus, the chunk parameter creates a trade off between communication overhead and load balance. Pipes architecture visualizations are presented at Figure~\ref{fig:pipe_configurations}. 

\subsection{Thread schedulers}
In OpenMP, the schedule clause controls how loop iterations are partitioned into chunks and assigned to the threads of a team \cite{muller25, OpenMP52Spec, thoman12}. 

\begin{itemize}
	\item With \textbf{static scheduling}, iterations are pre-divided into fixed-size chunks and assigned to threads in round-robin order; if no chunk size is given, the iteration space is split into approximately equal-sized chunks, usually minimizing scheduling overhead and working well for regular, balanced loops \cite{muller25, OpenMP52Spec}.
	\item With \textbf{dynamic scheduling}, each thread executes one chunk and then requests another until no work remains; if no static chunk size is provided (making it chunk-based), it defaults to 1, which improves load balance for irregular workloads at the cost of higher runtime overhead \cite{muller25, OpenMP52Spec}.
	\item With \textbf{guided scheduling}, threads also request chunks, but dynamically. Still, the chunk size starts larger and decreases as the number of unassigned iterations shrinks, assigned to the thread that finishes first, again defaulting to one when unspecified. This provides a compromise between the low overhead of static scheduling and the better balance of dynamic scheduling \cite{OpenMP52Spec, thoman12}.
\end{itemize}

In chunk-based parallel scheduling, the main scheduling policy may be static or dynamic, while the iteration space or task pool is partitioned into chunks, which are assigned to worker threads or processes; the chunk size may be fixed or adaptively adjusted at runtime based on observed workload behavior \cite{agathos22, booth22}. In chunk-stealing policies, tasks are pre-assigned to threads using static allocation \cite{blumofe1999}. However, threads that finish first steal tasks from the working queues of other threads, mostly in a blind round-robin manner \cite{clet-ortega14, guo10}. 

Adaptive schedulers focus on system attributes and thread-based performance metrics, followed by a probing mechanism that determines the assigned task batches per thread (chunk size) or even increases the number of threads (contention increase) assigned to a specific workload \cite{akturk2019, kontogiannis11, Mohtasham15}.

\subsection{AIMD scheduler}\label{s:2_3}

 In this paper, the Additive Increase Multiplicative Decrease (AIMD) adaptive scheduler is introduced, following the TCP protocol's congestion control mechanism \cite{jacobson1988, chiu1989, kontogiannis05}. This scheduler is an adaptive contention-aware window scheduler that controls the number of concurrently active work chunks using an additive-increase, multiplicative-decrease policy. A fixed pool of worker threads is created once, up to the maximum number of available threads or processors, and these workers remain alive throughout the execution. However, the scheduler does not allow all workers to issue work at all times. Instead, it maintains a contention window that limits the number of chunks that may be in flight simultaneously. 

The AIMD policy initializes the contention window to a small bootstrap value and then gradually adjusts parallelism by assigning additional chunks as long as the real-time measured EWMA (Exponential Weight Moving Average) raw utilization remains below the configured overload threshold. 

For each CPU core i, the probing monitor master process maintains a long-term arithmetic mean utilization over the completed sampling second intervals, updating it at each sample p as per core utilization U\_i (p), which is the current p-probed utilization percentage of core i. This produces one persistent mean-utilization value per i core.
In Linux, per-core utilization is usually calculated from the cumulative CPU-time counters exported in /proc/stat, specifically from the cpuN lines such as cpu0, cpu1, and so on. These counters are measured in 1/USER\_HZ=1/100Hz jiffies (ms=10ms) since boot and include fields such as user, system, idle, and iowait.
 
To obtain the current utilization of core i at probe p, a monitor reads the cpu\_i counters at two consecutive sampling instants, computes the delta for each field, sums the deltas to obtain the total elapsed CPU time, and then divides the busy delta time by that total. 
At the global many-core level, the monitor builds a combined global utilization metric based on Equation~\eqref{eq:utilization_combined}.

\begin{equation}
	\text{Ug}_p = a \cdot \max_i(\bar{U}_i) + (1 - a) \cdot \min_i(U_i(p))
	\label{eq:utilization_combined}
\end{equation}

This way it blends the most persistently busy core with the least busy core at the system level. The global utilization is then smoothed with an exponentially weighted moving average according to Equation~\eqref{eq:ewma}.

\begin{equation}
	U(p) = \beta \bar{U}_p + (1 - \beta) \text{Ug}_p
	\label{eq:ewma}
\end{equation}

with the first valid sample initializing the EWMA directly to the combined value. With the chosen parameters $a=0.7$ and $\beta=0.8$, the utilization metric gives strong-er weight to the maximum long-term per-core load than to the least-loaded current core, while also applying relatively strong smoothing, resulting in a stable, less sensitive to transient loads response that remains mainly responsive to sustained contention changes. 

In this AIMD-based scheduler, the U(p) utilization serves as an immediate feedback signal to deter-mine how many task chunks may be as-signed concurrently to worker threads. If the measured U(p) utilization exceeds a configured overload threshold, the scheduler interprets this as oversubscription or excessive contention and halves the contention window. Otherwise, it applies an additive increase, incrementing the window by one thread. 

As a slow-start mechanism in AIMD, when the U(p) function from Equation~\eqref{eq:ewma} equals zero, the contention window doubles at each chunk completion up to the processor count N, allowing it to ramp up quickly from an idle state. In this way, the scheduler seeks a stable operating point that maximizes useful concurrency while preventing overload due to excessive simultaneous chunk execution.

\subsection{Adaptive chunk scheduler}\label{s:2_4}
To outperform the guided thread scheduler, this paper presents an adaptive chunked scheduler. Its purpose is to determine the size of each assigned chunk by combining guided-style work partitioning with runtime observations of per-thread execution speed. At the beginning, no thread has a timing history, so every thread starts with speed\_factor \= 1.0, and the initial chunk is computed only from the remaining tasks as base\_chunk = remaining / k, which makes the scheduler initially behave like a guided scheduler 
\cite{kontogiannis2011probing,kontogiannis2014albl}. 

After a thread completes a chunk, its execution time is measured, and the scheduler stores the total time per thread in milliseconds and the total number of chunks finished per thread. The thread's average time per slice is then computed. It also computes a global average slice time across all threads with available history (global\_avg) and then defines the per-thread speed\_factor = global\_avg / thread\_avg. 

As a result, faster-than-average threads obtain $speed\_factor > 1$ and therefore receive larger chunks. In comparison, slower-than-average threads obtain $speed\_factor < 1$ and receive smaller chunks, according to Equation (\ref{eq:adaptive_chunk}).

\begin{equation}
	\mathit{chunk} = \mathit{base\_chunk} \cdot \mathit{speed\_factor}
	\label{eq:adaptive_chunk}
\end{equation}

with the final chunk always clamped so that $1 \le \mathit{chunk} \le \mathit{remaining}$. Consequently, this adaptive scheduler starts like guided scheduling, adapts the chunk size based on observed thread speed, and naturally reduces chunk sizes again near the end of execution as the remaining work becomes small, thus preserving lower scheduling overhead than fully dynamic scheduling while improving load balance when thread performance is heterogeneous.

\section{Metrics used}\label{s:3}

\subsection{Performance and Efficiency}\label{s:3_1}

For a fixed workload, the performance $P$ of an executing workload is defined as \textit{the inverse of the execution} time $T$. So, lower runtime for the scheduled workload implies higher performance, $P = 1/T$, for the implemented policy. Over a set of $n$ repeated executions, the maximum observed performance is denoted by $P_{max} = \max_{i}(P_i)$, and the normalized performance or efficiency is computed as $E_p$ with $0 \leq E_p \leq 1$. In this context, an efficiency value of 0.8 across 10 or more experiment repetitions can be considered a good, stable performance level, as it indicates that the measured execution remains close to the best observed run.

\subsection{Speedup and SEfficiency}\label{s:3_2}
The speedup of a parallel system, denoted by $\sigma(N)$, is defined as the ratio of the execution time of the serial version to the execution time with $N$ processors, according to Equation (\ref{eq:sigma_speedup}).

\begin{equation}
	\sigma(N) = \frac{T_{serial}}{T_{parallel}(N)}
	\label{eq:sigma_speedup}
\end{equation}

Equation (\ref{eq:sigma_speedup}) measures how much faster the parallel implementation of a workload on $N$ processors is compared to the sequential one. Based on speedup, the parallel speedup efficiency ($\sigma Efficiency$) is defined using Equation (\ref{eq:se_efficiency}).

\begin{equation}
	\sigma E(N) = \frac{\sigma(N)}{N}
	\label{eq:se_efficiency}
\end{equation}

Equation (\ref{eq:se_efficiency}) expresses the effectiveness with which the available processors are utilized. An efficiency close to 1 indicates near-ideal processor utilization, while lower values indicate increasing overhead due to communication, synchronization, or load-core imbalance. In practice, $\sigma Efficiency$ values around 0.8 or higher are usually considered very good; values near 0.6 remain acceptable. However, values below 0.4 indicate limited benefit from further parallelization.

For a system with $N$ fixed cores, a parallel code is considered to scale well when scheduling $k=1..K$ threads or processes leads to increasing speedup $\sigma(k)$ and sufficiently high $\sigma Efficiency$ (Speedup Efficiency). To provide forced core assignment per $k$-thread, processor affinity may be forced in a round-robin manner (if $K > N$). In practice, good scalability is observed when the gain from increasing $k$ remains significant, and efficiency stays above about 0.6, with values near 0.8 indicating very good resource utilization.

\subsection{Amdahl law and the effective serialization coefficient}\label{s:3_3}

Amdahl's law \cite{Amdahl1967} states that the speedup of a parallel program is fundamentally limited by the fraction of the code that must remain executing in a serial manner. If $\alpha$ denotes the serial portion of mutex synchronizations and atomic operations on the workload and $1-\alpha$ the parallelizable portion, then the theoretical speedup on $k=N$ processors is estimated using Equation (\ref{eq:amdahl_sigma}).

\begin{equation}
	\sigma(N) = \frac{1}{\alpha + \frac{1-\alpha}{N}}
	\label{eq:amdahl_sigma}
\end{equation}

This means that even if the number of $N$ processors increases significantly, the maximum achievable speedup cannot exceed the bound imposed by the serial part of the program. Therefore, Amdahl's law is widely used to evaluate the scalability limits of parallel codes, and schedulers boosts to estimate how much performance improvement can realistically be obtained from additional computing resources.

When estimating the serial fraction $\alpha$ from Amdahl's law, the efficiency relation can be written according to Equation (\ref{eq:se_alpha}).

\begin{equation}
	\sigma E(k) = \frac{\sigma(k)}{k} = \frac{1}{1 + \alpha(k-1)}
	\label{eq:se_alpha}
\end{equation}

which gives the estimate for $\alpha$ using Equation (\ref{eq:alpha_estimate}).

\begin{equation}
	\alpha = \frac{\frac{1}{\sigma E(k)} - 1}{k-1} = \frac{1 - \sigma E(k)}{\sigma E(k)(k-1)}
	\label{eq:alpha_estimate}
\end{equation}

However, if the number of scheduled threads or processes $k$ exceeds the number of available cores $N$, then $k$ no longer represents true hardware parallelism (unless affinity is forced), because oversubscription and excessive processing delays introduce context-switching and synchronization overhead. In that case, a better approximation is to use the effective parallelism, $k_{eff} \approx N$ (more generally, thread affinity applies in a round robin manner and $k_{eff} = \min(k, N)$). Then the estimate of the serial fraction from the $\sigma Efficiency$ is measured at the hardware limit or hardware overutilization. If $\sigma E(k_{eff})$ is close to 0.8, indicating a relatively small serial fraction and therefore good parallel scalability up to the available $N$ cores or even above to the $k_{eff}$ if $k_{eff} > N$. The $\alpha$ estimate in that case is better than the $\alpha$ at $k = N$.

For consistency, all $\sigma E(k)$ and $\alpha(k)$ values reported in 
the tables of Section~6 are computed using the nominal worker count 
$k$ rather than the hardware-effective $k_{eff}$, even when $k > N$. 
This convention reflects software-level parallelism as configured by 
the user, and intentionally exposes the efficiency collapse caused by 
oversubscription. The $k_{eff}$ formulation introduced above should be 
understood as an alternative interpretive lens rather than the 
convention applied to the tabulated results.

By construction, the serial fraction $\alpha$ in Amdahl's law is 
bounded in $[0,1]$. The estimator $\alpha=(1-\sigma E(k))/(\sigma E(k)(k-1))$ 
used here is the corresponding Karp--Flatt diagnostic \cite{KarpFlatt1990} and 
is defined only for $k \geq 2$. In practice, execution-time variance 
introduced by operating system interruptions and CPU scheduler 
decisions occasionally pushes the measured efficiency marginally 
above $1.0$, producing small negative values of $\alpha$ on the order 
of $10^{-4}$; these do not indicate super-linear speedup but rather 
normal measurement noise. Conversely, when the measured speedup 
satisfies $\sigma(k) < 1$ (parallel slowdown, typically from 
oversubscription overhead exceeding the parallelizable gain), the 
estimator returns formal values $\alpha > 1$. In both cases the 
Amdahl decomposition does not hold, and $\alpha$ should be read as a 
diagnostic indicator rather than as a meaningful serial fraction; 
the corresponding rows are retained for transparency.

\subsection{Amdahl vs Gustafson serialization coefficient}\label{s:3_4}

Under Amdahl's law, when the workload is kept fixed, even a small serial fraction can strongly limit scalability; for example, with $N=24$ and $\alpha=0.05$ (5\% serializable code), the speedup remains far below the ideal value of 24 because the 5\% of serial code becomes a bottleneck. In contrast, under Gustafson's formulation \cite{Gustafson88}, as also mentioned in Equation (\ref{eq:gustafson}).

\begin{equation}
	\sigma(N) = N - \alpha_{scaled}(N-1)
	\label{eq:gustafson}
\end{equation}

Based on Equation (\ref{eq:gustafson}) when execution time is approximately constant, and the problem size grows proportional to the number of processors, the parallel portion increases, and the speedup can remain close to $N$. Therefore, Amdahl's law is the appropriate model for fixed-size workloads, whereas Gustafson's law better captures scalable parallel computing $\alpha$ values ($\alpha_{scaled}$), when larger problems are solved with more processors. In our experiments, we use a fixed-size workload and test different workload-balancing schedulers \cite{Amdahl1967, Gustafson88}. Therefore, the serial portion $\alpha$ from Amdahl's law is different from Gustafson's $\alpha$. In Amdahl's law, $\alpha$ usually represents the serial fraction of the original workload-code. In contrast, in Gustafson's law, $\alpha$ means the serial fraction of the execution time on the parallel system.

\subsection{Karp-Flatt serialization coefficient}\label{s:3_5}

For the scheduled threads or processes on $N$ physical cores, the Karp–Flatt metric can be used to detect or indicate scaling collapse \cite{KarpFlatt1990}. 
The Karp–Flatt metric (effective serial fraction) can be used as a practical indicator of scaling collapse, provided it is interpreted as an apparent rather than purely algorithmic serial fraction.

Instead of using the oversubscribed software parallelism $k$, one uses the effective hardware parallelism, $k_{eff} = \min(k, N) \approx N$, and computes the Karp–Flatt metric according to Equation (\ref{eq:karp_flatt}).

\begin{equation}
	\varepsilon(k_{eff}) = \frac{\frac{1}{\sigma(k_{eff})} - \frac{1}{k_{eff}}}{1 - \frac{1}{k_{eff}}} \approx \alpha
	\label{eq:karp_flatt}
\end{equation}

This $\varepsilon(k_{eff})$ is an approximation for $k=N$ to the serial part of the executed workload. If this apparent Karp–Flatt fraction increases as $k$ grows beyond $N$, then the program is no longer gaining useful parallelism and is instead drowning in overhead, such as context switching, synchronization, scheduling pressure, and load imbalance. This adapted formula is useful for detecting when extra software concurrency stops helping and starts hurting scalability.

\subsection{Utilization}\label{s:3_6}

Utilization measures the fraction of the available execution time during which a workload effectively engages a resource. For each task assigned to a thread, the thread utilization can be defined and measured as the fraction of the thread's observed lifetime during which it is engaged in useful activity. In practice, useful activity includes the user's execution time ($T_{user}$), system/kernel execution time ($T_{system}$), and I/O wait time ($T_{iowait}$), while the time spent idle, either sleeping or waiting in a queue, is moved to the denominator and treated as non-productive delay. Thus, the utilization of a thread executing a task is expressed using Equation (\ref{eq:u_thread}).

\begin{equation}
	U_{thread} = \frac{T_{user} + T_{system} + T_{iowait}}{T_{user} + T_{system} + T_{iowait} + T_{idle}}
	\label{eq:u_thread}
\end{equation}

where $T_{idle}$ represents the total time the thread remains inactive due to sleep state or queue waiting. Therefore, higher utilization values indicate that the assigned task keeps the thread busy for a larger fraction of its lifetime. In contrast, lower values reveal increasing idle overhead and poorer task-to-thread allocation. This utilization metric is used by the AIMD task scheduler to assign, in real time, different batches of workload chunks.

\subsection{Many-Core System Utilization}\label{s:3_7}

For a many-core system with $N$ cores, the overall CPU utilization can be written as the ratio of total active core time to total available core time, using Equation (\ref{eq:u_n_simple}).

\begin{equation}
	U_N = \frac{\sum_{i=1}^{N} T_{active,i}}{N \cdot T_{total}}
	\label{eq:u_n_simple}
\end{equation}

expressed using $M$ processes containing threads on an $N$-core system:

\begin{equation}
	U_N = \frac{\sum_{i=1}^{M} (T_{user,i} + T_{system,i} + T_{iowait,i})}{N \cdot T_{total}}
	\label{eq:u_n_processes}
\end{equation}

That is, Equation~\eqref{eq:u_n_processes} can be rewritten as:

\begin{equation}
	U_N = \frac{\sum_{i=1}^{M} (T_{user,i} + T_{system,i} + T_{iowait,i})}{N \left( \max_{1 \le i \le M} t_i^{end} - \min_{1 \le i \le M} t_i^{start} \right)}
	\label{eq:u_n_final}
\end{equation}

where $t_i^{start}$ and $t_i^{end}$ denote the start and end times of process $i$, respectively.

\subsection{Parallel Overhead Time}\label{s:3_8}

Let $T_1$ the execution time on 1 core, $T_k$ the execution time using $k$ threads, with $k \le N$, $T_p$ the parallelizable code part as executed in one single core, $T_p/k$ the execution time on $k$ threads of the parallelizable code part, $T_s$ the total execution time of the serial portion of the code, the intrinsic serial fraction of the workload $\alpha$, $1-\alpha$ the parallelizable fraction, then, the total parallel overhead $T_{ov}(k)$ using $k$ threads can be calculated as follows:

\begin{equation}
	T_1 = T_s + T_p, \qquad T_k = T_s + \frac{T_p}{k} + T_{ov}(k)
	\label{eq:overhead_decomp}
\end{equation}

with $T_s = \alpha T_1$, $T_p = (1-\alpha)T_1$ gives Equation (\ref{eq:overhead_final}).

\begin{equation}
	T_{ov}(k) = T_k - \left( \alpha T_1 + \frac{(1-\alpha)T_1}{k} \right)
	\label{eq:overhead_final}
\end{equation}

Using the measured speedup, the overhead can also be written as:

\begin{equation}
	f(k) = \frac{T_{ov}(k)}{T_1} = \frac{1}{\sigma(k)} - \alpha - \frac{1-\alpha}{k}
	\label{eq:overhead_ratio}
\end{equation}

Moreover, if $\varepsilon(k) \ne \alpha$ is the Karp–Flatt metric, then:

\begin{equation}
	T_{ov}(k) = T_1 \left( 1 - \frac{1}{k} \right) (\varepsilon(k) - \alpha)
	\label{eq:overhead_kf}
\end{equation}

\subsection{Distributed Systems Amdahl's law}\label{s:3_9}

In a distributed system, Amdahl's law models the speedup by separating the workload into a serial fraction $\alpha$ and a parallel fraction $1-\alpha$.

Suppose the system consists of $P$ distributed nodes, and each node may maintain $k$ processes or threads through the MPI-based execution model, giving a total of $P_k = P \times k$ concurrent execution units. Under this assumption, the ideal speedup can be expressed using Equation~\eqref{eq:dist_amdahl}.

\begin{equation}
	\sigma(P_k) = \frac{1}{\alpha + \frac{1-\alpha}{P_k}}
	\label{eq:dist_amdahl}
\end{equation}

In practice, however, in distributed environments, $\alpha$ includes not only strictly sequential computation, but also communication overhead, synchronization delays, message passing, data movement, and load imbalance among MPI processes or threads. If $T_1$ is the sequential execution time and $T_{P_k}$ is the execution time using $P$ nodes with $k$ MPI processes or threads per node, then the measured speedup is $\sigma(P_k) = \frac{T_1}{T_{P_k}}$. Rearranging Amdahl's law gives the effective serial fraction using Equation~\eqref{eq:dist_alpha}.

\begin{equation}
	\alpha = \frac{\frac{1}{\sigma(P_k)} \cdot P_k - 1}{P_k - 1} = \frac{\frac{P_k}{\sigma(P_k)} - 1}{P_k - 1}
	\label{eq:dist_alpha}
\end{equation}

This empirical estimate is closely related to the Karp–Flatt metric, which is calculated using $k=1$ in Equation~\eqref{eq:overhead_kf}, and is particularly useful in distributed systems because it captures the overall non-scalable part of the execution. A small $\alpha$ indicates efficient parallel execution, whereas an increasing Karp–Flatt $\alpha$ as the number of nodes, processes, or threads grows suggests that communication and coordination overheads increasingly limit scalability.

\subsection{Universal Scalability Law (USL-Gunther)}\label{s:3_10}

For distributed systems, a more realistic scalability model than Amdahl’s law is Gunther's Universal Scalability Law (USL) \cite{gunther1993usl, gunther_general_2008}, since it accounts not only for serialization effects due to racing conditions in critical sections or segments, but also for contention and communication or coherence overheads separately from serialization. If the distributed platform consists of $P$ nodes and each node maintains $k$ MPI processes or threads \cite{gropp1999usingmpi, mpi41standard}, then the total number of parallel execution units is $N = P \times k$. The measured speedup is defined as $\sigma(N) = \frac{T_1}{T_N}$, where $T_1$ is the sequential execution time and $T_N$ is the execution time with $N$ workers. According to USL, the speedup is modeled using Equation~\eqref{eq:usl_sigma}.

\begin{equation}
	\sigma(N) = \frac{N}{1 + \alpha(N-1) + \beta N(N-1)}
	\label{eq:usl_sigma}
\end{equation}

where $\alpha$ represents the contention synchronization cost for shared resources, and $\beta$ represents the coherence or communication cost among processes/nodes. To estimate these parameters from experimental data, one defines:

\begin{equation}
	Y(N) = \frac{N}{\sigma(N)} - 1
	\label{eq:usl_y}
\end{equation}

which leads to the linear form of Equation~\eqref{eq:usl_linear}.

\begin{equation}
	\frac{Y(N)}{N-1} = \alpha + \beta N = \frac{\frac{N}{\sigma(N)} - 1}{N-1}
	\label{eq:usl_linear}
\end{equation}

Thus, using speedup measurements for different values of $N$, the parameters $\alpha$ and $\beta$ can be obtained either from two measurements or more reliably through linear regression. 

When $\beta = 0$, the USL reduces to Amdahl’s law, while when both $\alpha \approx 0$ and $\beta \approx 0$, the system approaches ideal linear scaling. Finally, the distributed overhead time relative to ideal execution is given by Equation~\eqref{eq:usl_overhead}.

\begin{equation}
	\begin{aligned}
		T_{over}(N)
		&= T_N - \frac{T_1}{N} \\
		&= \frac{T_1}{N}
		\left[
		\alpha(N-1) + \beta N(N-1)
		\right]
	\end{aligned}
	\label{eq:usl_overhead}
\end{equation}

Therefore in cases where  $\beta > 0$, the speedup for $N$ processes is calculated Equation~\eqref{eq:usl_sigma}.
\section{Implementation Details}\label{s:4}
All schedulers were implemented in C++ and evaluated on the same row-sorting workload. The 
input data consist of a three-dimensional tensor of unsigned 8-bit values with dimensions \(F 
\times H \times W\), where \(F\) denotes the number of frames and each frame contains \(H\) 
rows of length \(W\). In the experiments, the workload uses tensors of size \(1000 \times 640 
\times 640\) and \(10000 \times 640 \times 640\).

Each task corresponds to one frame, and the worker assigned to that task sorts all rows of that frame independently using quick sort. 
To ensure that all schedulers operate on identical input data, the original tensor is generated once and stored in a System V shared-memory segment. Before each scheduler execution, a 
separate shared-memory work copy of the original tensor is created. 
This prevents later schedulers from operating on data already sorted by earlier runs while preserving identical initial 
input across schedulers. 

Although the benchmark executes the schedulers sequentially for measurement isolation, each scheduler internally processes its assigned workload in parallel using multiple worker processes 
or threads. The use of separate shared-memory copies guarantees that every scheduler operates on the same initial unsorted tensor while avoiding interference between consecutive 
experimental runs. 
The tensor is stored in row-major order using the indexing formula 
$$A[(fH+i)W+j],$$ 

where \(f\) is the frame index, \(i\) is the row index, and \(j\) is the column index. During sorting, 
each row is wrapped by a lightweight row view and sorted directly inside the corresponding 
memory segment without deep-copying the row data. For process-based schedulers, worker processes are created using \texttt{fork()}. The bounded 
prolific scheduler creates all workers directly from the parent process and assigns frames 
cyclically, so worker \(w\) executes tasks \(w, w+k, w+2k, \ldots\). CPU (Central Processing Unit) affinity can optionally be 
enabled, assigning workers to CPUs in a round-robin manner using \texttt{sched\_setaffinity()}. Thread schedulers were implemented using Boost threads and shared-memory task distribution structures. 
Depending on the scheduling policy, workloads were assigned dynamically, through 
fixed-size chunks, guided chunk allocation, adaptive chunk resizing, or work-stealing mechanisms. All schedulers operated on the same frame-level row-sorting workload and shared-memory tensor representation.
The AIMD scheduler additionally maintains a contention window initialized to $\min(k, N)$, processes frames in chunks of size $\mathrm{CHUNK} = 4$, halves the window whenever the EWMA utilization exceeds $95\%$, and reads the EWMA value through a lock-free atomic mirror sampled once every $32$ chunk completions. 

For every scheduler family, the measured execution time begins immediately before the scheduler's entry point is invoked and ends when the last worker completes its assigned work. Consequently, for process-based schedulers (bounded prolific, bounded collective), the measurement includes all \texttt{fork()} calls and child-process startup; for pipe-based schedulers, it includes pipe creation and the first read/write handshake;and for thread-based schedulers, it includes the creation and startup of the worker-thread pool. Synchronization primitives (mutexes, condition variables, barriers) used during task distribution are likewise included in the measured interval for all three families. One exception applies to the AIMD scheduler: its CPU-utilization monitor is instantiated once before any timed run and shared across repeated executions, so its one-time startup latency is excluded from the reported AIMD runtimes; this avoids penalizing AIMD for a cost unrelated to its scheduling logic. Excluded from all measurements are only the one-time steps that are common to every scheduler and unrelated to the scheduling policy itself: generation of the original tensor and 
allocation of the shared-memory segments. After each run, the output tensor is verified to ensure that every row is sorted in nondecreasing order. 


\section{Experimental Scenario}\label{s:5}
\label{sec:experimental_scenario}

The experiments were conducted on a server equipped with four Intel Xeon X7460 processors operating at 2.66 GHz, providing a total of 24 physical cores (24 logical processing units). The system was configured with 62 GB of RAM and ran Ubuntu 18.04.6 LTS (Bionic Beaver) with Linux kernel version 4.19.29. 

All benchmark applications were compiled using G++ (GCC) version 7.5.0 with the compiler flags `-std=c++14 -Wall -Wextra -O3 -fpermissive`, and linked against the Boost.Thread, Boost.System, Boost.Chrono, and POSIX Threads libraries. The experiments were performed on an otherwise idle system to minimize interference from background processes. No explicit CPU affinity was enforced for either processes or threads. Worker placement was managed by the default Linux scheduler, and no CPU pinning or NUMA-aware memory placement techniques were employed.

The evaluated workload consists of row-wise sorting operations on shared-memory tensors of dimensions $1000 \times 640 \times 640$ and $10000 \times 640 \times 640$, corresponding approximately to 0.38 GB and 3.81 GB of unsigned 8-bit data, respectively. Although the full memory footprint includes both the original tensor and its work copy (totalling 0.76 GB and 7.64 GB respectively), the reported sizes reflect only the work copy operated on by each scheduler, since the original tensor remains read-only throughout execution.

Each experiment was repeated 10 times to ensure statistical reliability and to account for operating system jitter. The reported execution metrics, including execution time and speedup, were computed using the arithmetic mean of the measured execution times across these 10 repetitions. Using the average execution time provides a realistic representation of the scheduler's expected performance under typical system conditions. Furthermore, a strict performance threshold of 1\% was applied during the comparative analysis. Execution time or speedup differences smaller than 1\% fall within the margin of unavoidable environmental noise. Consequently, schedulers with performance differences below this threshold are treated as practically equivalent in those specific configurations. 

The experiments evaluate both process-based and thread-based schedulers under different levels of parallelism using varying numbers of worker processes or threads $k$. 
Oversubscription refers to configurations in which the number of worker entities exceeds the available logical processing units of the system. 
Measurements include execution time, speedup $\sigma(k)$, speedup efficiency $\sigma_E(k)$, and the estimated serialization 
factor $\alpha(k)$. 
Unless otherwise stated, processor affinity was enabled using 
\texttt{sched\_setaffinity()} in order to study the impact of explicit CPU binding on 
scheduler scalability, processor contention, and execution stability. 

For every scheduler family, the measured execution time begins immediately 
before the scheduler's entry point is invoked and ends when the last worker 
completes its assigned work. Consequently, for process-based schedulers 
(bounded prolific, bounded collective), the measurement includes all 
\texttt{fork()} calls and child-process startup; for pipe-based schedulers, 
it includes pipe creation and the first read/write handshake; and for 
thread-based schedulers, it includes the creation and startup of the 
worker-thread pool. Synchronization primitives (mutexes, condition 
variables, barriers) used during task distribution are likewise included 
in the measured interval for all three families. One exception applies to 
the AIMD scheduler: its CPU-utilization monitor is instantiated once before 
any timed run and shared across repeated executions, so its one-time 
startup latency is excluded from the reported AIMD runtimes; this avoids 
penalizing AIMD for a cost unrelated to its scheduling logic. Excluded 
from all measurements are only the one-time steps that are common to 
every scheduler and unrelated to the scheduling policy itself: generation 
of the original tensor and allocation of the shared-memory segments.

This measurement methodology was adopted to ensure that the reported results reflect not only the computational cost of the row-wise sorting workload, but also the practical overhead introduced by each scheduling strategy. Therefore, the comparison captures the end-to-end behavior of each scheduler under realistic execution conditions, including worker creation, task dispatching, inter-process or inter-thread coordination, synchronization, and load-balancing decisions.

\subsection{Scenario I. Process parallelism}\label{s:5_1}

\begin{figure}[ht]
	\centering
	\includegraphics[width=0.48\textwidth, height=6cm, keepaspectratio=false]{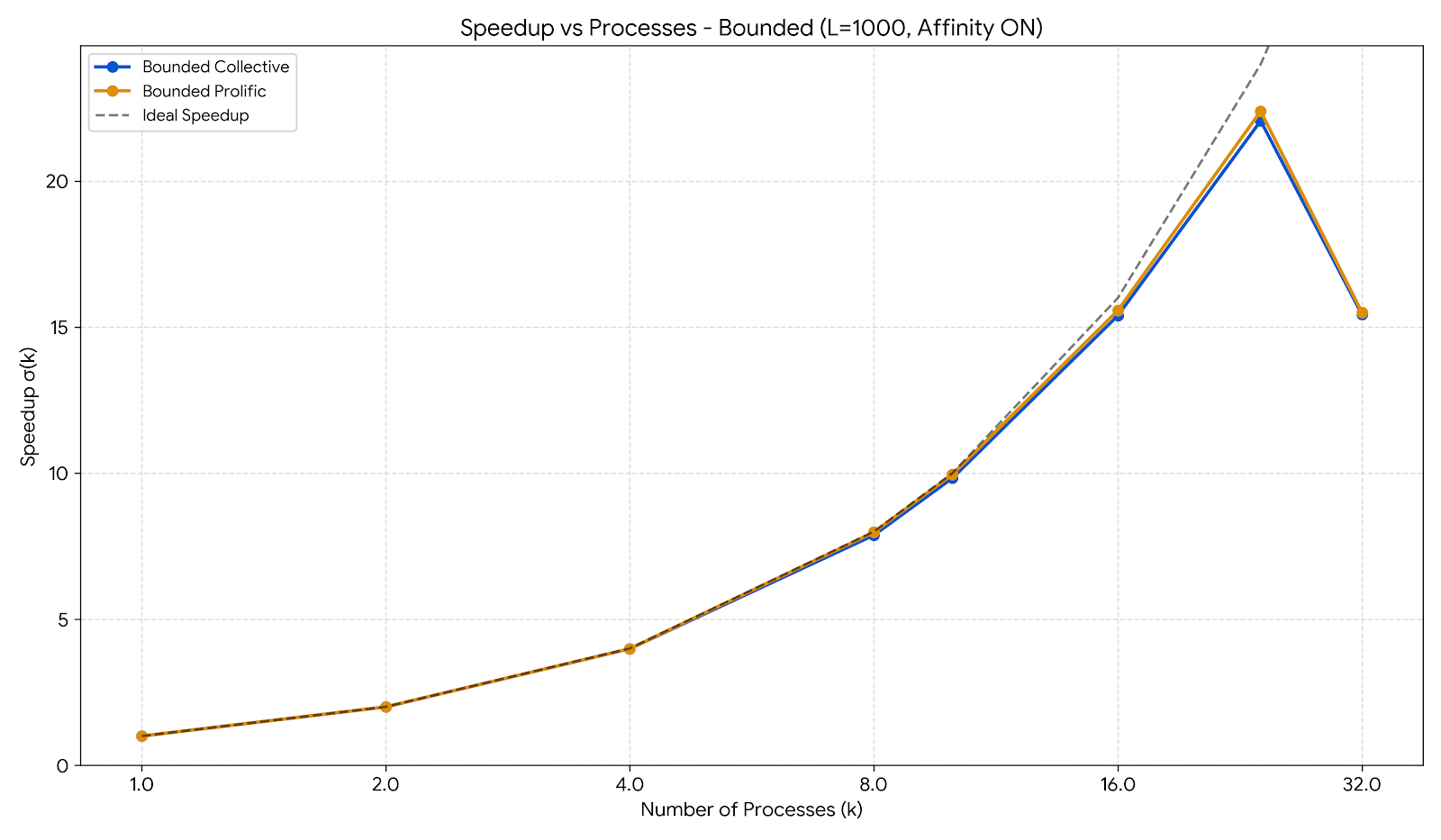}
	\caption{Performance and speedup metrics for the Bounded Prolific and Collective Schedulers using L=1000.}
	\label{fig:bounded_L1000}
\end{figure}

\begin{figure}[ht]
	\centering
	\includegraphics[width=0.48\textwidth, height=6cm, keepaspectratio=false]{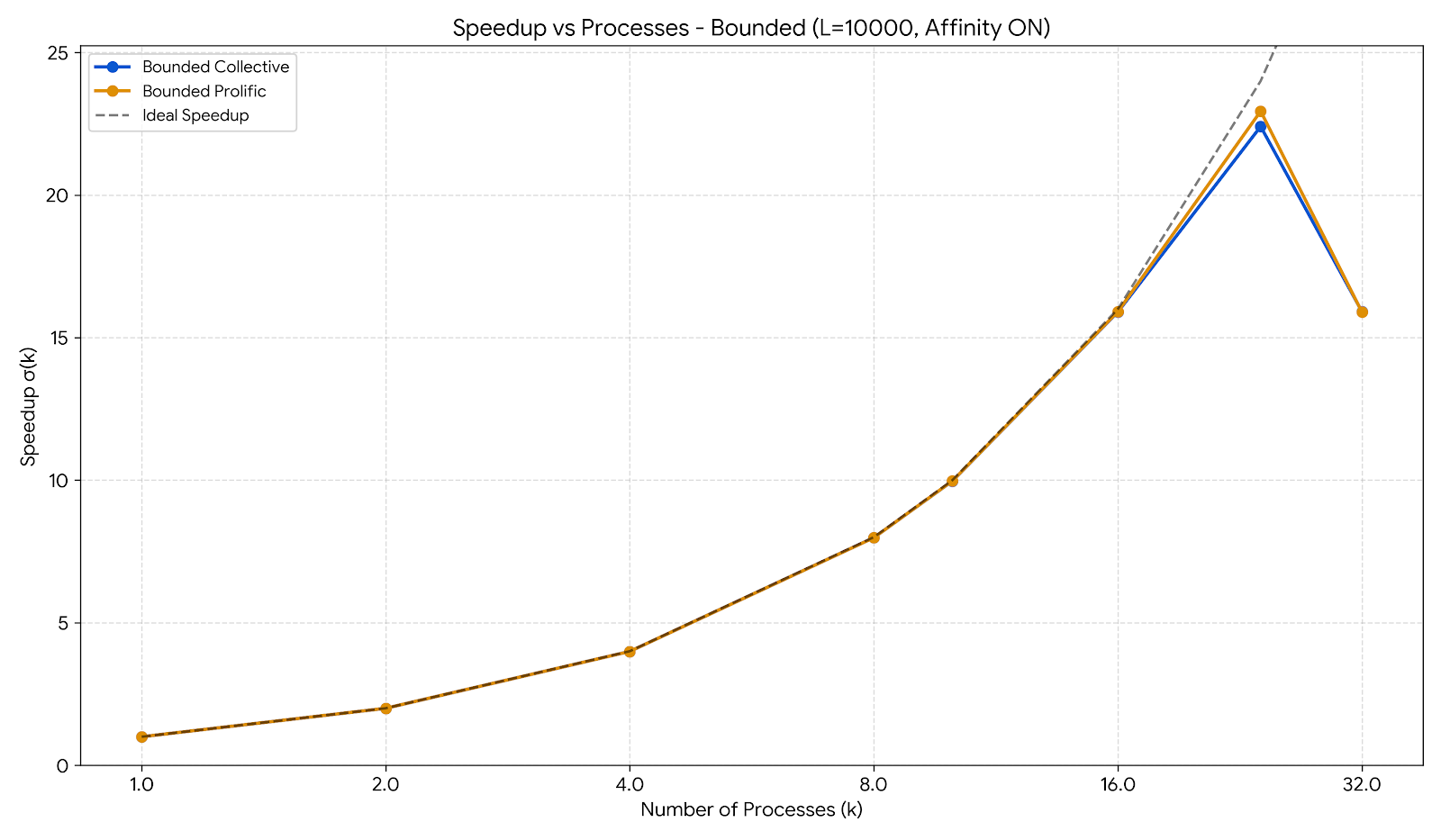}
	\caption{Performance and speedup metrics for the Bounded Prolific and Collective Schedulers using L=10000.}
	\label{fig:bounded_L10000}
\end{figure}

\begin{figure}[ht]
	\centering
	\includegraphics[width=0.48\textwidth, height=6cm, keepaspectratio=false]{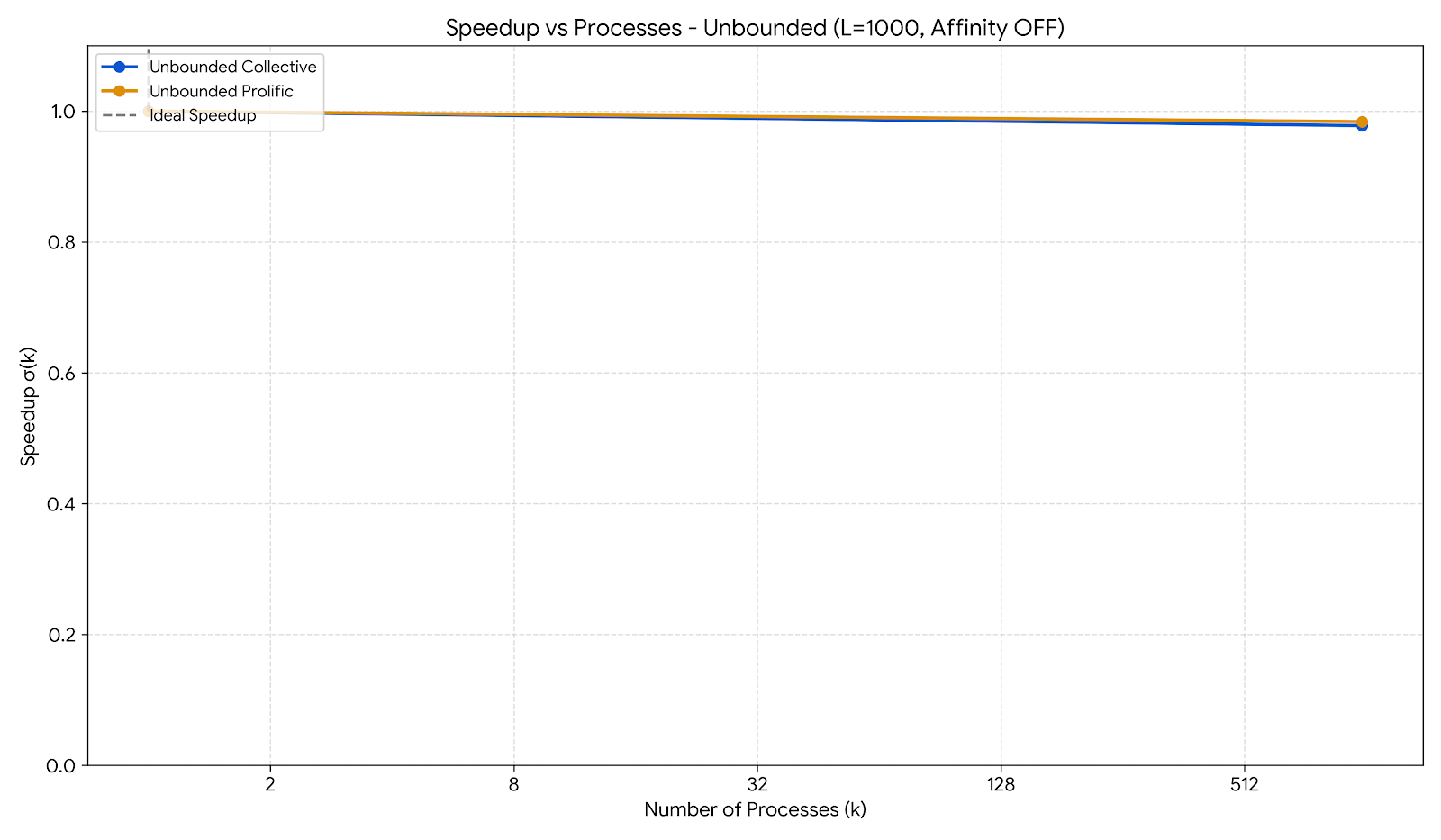}
	\caption{Performance and speedup metrics for the Unbounded Prolific and Collective Schedulers using L=1000.}
	\label{fig:unbounded_L1000}
\end{figure}

\begin{figure}[ht]
	\centering
	\includegraphics[width=0.48\textwidth, height=6cm, keepaspectratio=false]{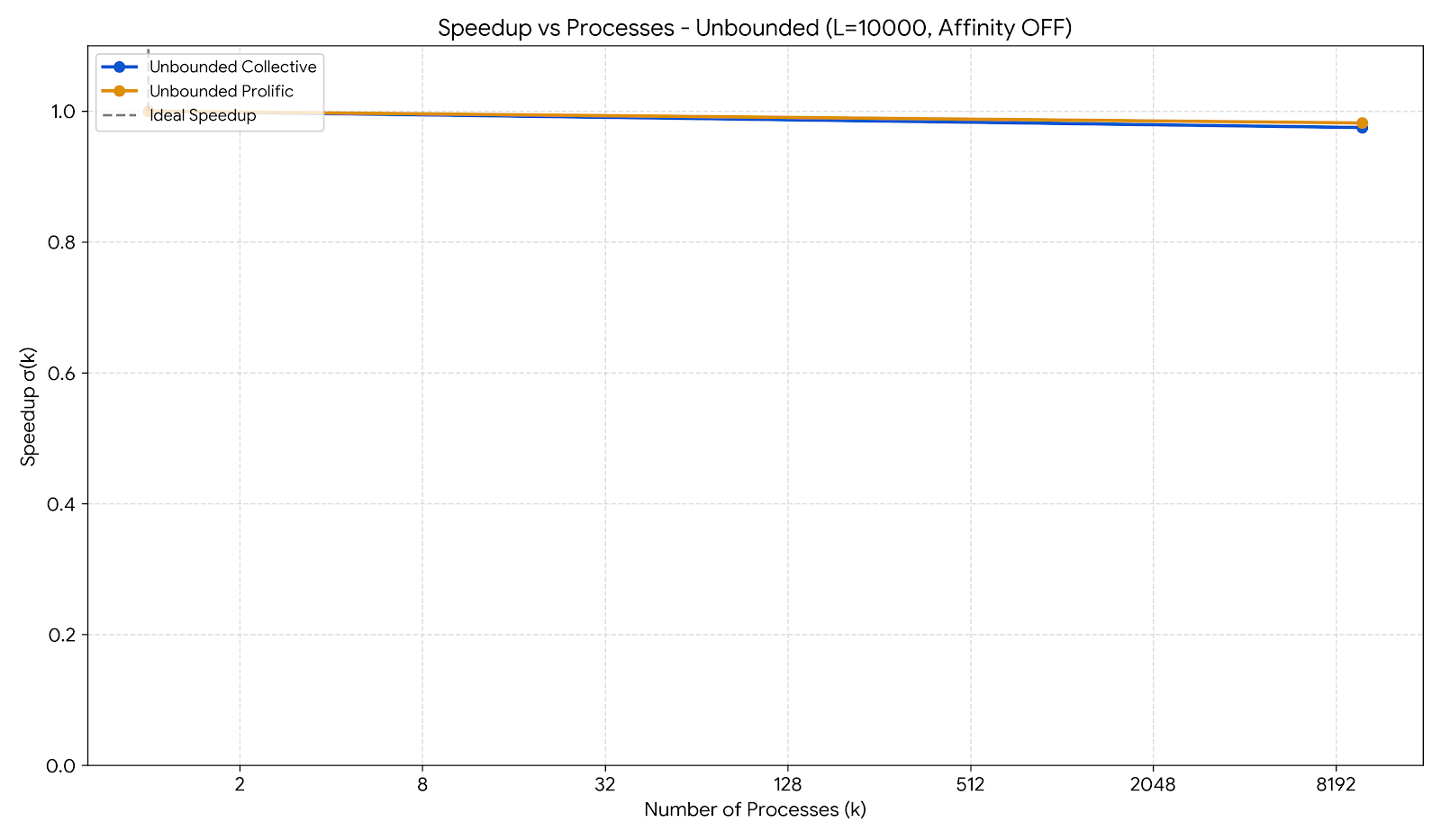}
	\caption{Performance and speedup metrics for the Unbounded Prolific and Collective Schedulers using L=10000.}
	\label{fig:unbounded_L10000}
\end{figure}

Process-based scheduling is evaluated for heavyweight tasks. Prolific forking increases contention additively, while collective forking uses a subreaper/leader model \cite{vasileiadou25}. Due to Copy-on-Write (COW), shared memory is essential for SIMP task allocation. The bounded prolific and bounded collective schedulers with processor affinity are presented in Tables~\ref{tbl:prolific_aff} and~\ref{tbl:collective_aff}, with corresponding speedup plots in Figures~\ref{fig:bounded_L1000} and~\ref{fig:bounded_L10000}. Their unbounded variants without affinity appear in Tables~\ref{tbl:unbounded_prolific} and~\ref{tbl:unbounded_collective}, with corresponding plots in Figures~\ref{fig:unbounded_L1000} and~\ref{fig:unbounded_L10000}. 

\begin{table}[ht]
	\centering
	\caption{Bounded prolific scheduler with affinity using L=1000,10000 ($640\times640$) 1-byte unsigned 2D tensors and row sorting using quick sort (N=24) }
	\label{tbl:prolific_aff}
	\begin{tabular}{|l|c|c|c|c|c|c|}
		\hline
		Scheduler & L& k & $\sigma(k)$ & $\sigma E(k)$ & $\alpha(k)$ \\ \hline
		\textbf{prolific} & \textbf{1000} & \textbf{2} & \textbf{1.999} & \textbf{0.999} & \textbf{0.001} \\ \hline
		prolific & 1000     & 4 & 3.992 & 0.998 & 0.001 \\ \hline
		\textbf{prolific} & \textbf{1000}     & \textbf{8} & \textbf{7.980} & \textbf{0.997} & \textbf{0.000} \\ \hline
		\textbf{prolific} & \textbf{1000}     & \textbf{10} & \textbf{9.961} & \textbf{0.996} & \textbf{0.000} \\ \hline
		\textbf{prolific} & \textbf{1000}     & \textbf{16} & \textbf{15.575} & \textbf{0.973} & \textbf{0.002} \\ \hline
		\textbf{prolific} & \textbf{1000}     & \textbf{24} & \textbf{22.394} & \textbf{0.933} & \textbf{0.003} \\ \hline
		\textbf{prolific} & \textbf{1000}     & \textbf{32} & \textbf{15.504} & \textbf{0.485} & \textbf{0.034} \\ \hline
		\textbf{prolific} & \textbf{10000}   & \textbf{2} & \textbf{1.999} & \textbf{0.999} & \textbf{0.001} \\ \hline
		\textbf{prolific} & \textbf{10000}  & \textbf{4} & \textbf{3.993} & \textbf{0.998} & \textbf{0.001} \\ \hline
		prolific & 10000  & 8 & 7.981 & 0.997 & 0.000 \\ \hline
		\textbf{prolific} & \textbf{10000} & \textbf{10} & \textbf{9.972} & \textbf{0.997} & \textbf{0.000} \\ \hline
		\textbf{prolific} & \textbf{10000}  & \textbf{16} & \textbf{15.914} & \textbf{0.995} & \textbf{0.000} \\ \hline
		\textbf{prolific} & \textbf{10000}  & \textbf{24} & \textbf{22.940} & \textbf{0.956} & \textbf{0.002} \\ \hline
		prolific & 10000  & 32 & 15.895 & 0.496 & 0.032 \\ \hline
	\end{tabular}
\end{table}
\begin{table}[ht]
	\centering
	\caption{Bounded collective scheduler with affinity using L=1000,10000 ($640\times640$) 1-byte unsigned 2D tensors and row sorting using quick sort (N=24) }
	\label{tbl:collective_aff}
	\begin{tabular}{|l|c|c|c|c|c|c|}
		\hline
		Scheduler & L & k & $\sigma(k)$ & $\sigma E(k)$ & $\alpha(k)$ \\ \hline
	\textbf{collective} & \textbf{1000} & \textbf{2} & \textbf{1.999} & \textbf{0.999} & \textbf{0.001} \\ \hline
	\textbf{collective} & \textbf{1000} & \textbf{4} &  \textbf{3.998} & \textbf{0.997} & \textbf{0.001} \\ \hline
	collective & 1000     & 8 & 7.877 & 0.985 & 0.002 \\ \hline
	collective & 1000     & 10 & 9.831  & 0.983  & 0.002 \\ \hline
	collective & 1000     & 16 & 15.398 & 0.962  & 0.003 \\ \hline
	collective & 1000     & 24 & 22.064 & 0.919 & 0.004 \\ \hline
	collective & 1000     & 32 & 15.431 & 0.482  & 0.035 \\ \hline
	collective & 10000   & 2 & 1.998 & 0.999 & 0.001\\ \hline
	\textbf{collective} & \textbf{10000}  & \textbf{4} & \textbf{3.993} & \textbf{0.998} & \textbf{0.001} \\ \hline
	\textbf{collective} & \textbf{10000}  & \textbf{8} & \textbf{7.983} & \textbf{0.998} & \textbf{0.000} \\ \hline
	collective & 10000  & 10 & 9.968 & 0.997 & 0.00 \\ \hline
	collective & 10000  & 16 & 15.899 & 0.994 & 0.00 \\ \hline
	collective & 10000  & 24 & 22.403 & 0.933 & 0.003 \\ \hline
	\textbf{collective} & \textbf{10000}  & \textbf{32} & \textbf{15.911} & \textbf{0.497} & \textbf{0.032} \\ \hline
	\end{tabular}
\end{table}


\begin{table}[ht]
	\centering
	\caption{Unbounded prolific scheduler using L=1000,10000 ($640\times640$) 1-byte unsigned 2D tensors and row sorting using quick sort (N=24) }
	\label{tbl:unbounded_prolific}
	\footnotesize
	\begin{tabular*}{\columnwidth}{@{\extracolsep{\fill}}|l|l|l|c|c|c|}
		\hline
		Scheduler & L & k & $\sigma(k)$ & $\sigma E(k)$ & $\alpha(k)$ \\ \hline
		\textbf{prolific} & \textbf{1000}     & \textbf{1000} & \textbf{0.984} & $\mathbf{9.84 \times 10^{-4}}$ & \textbf{1.00} \\ \hline
		\textbf{prolific} & \textbf{10000}  & \textbf{10000} & \textbf{0.982} & $\mathbf{9.80 \times 10^{-5}}$ & \textbf{1.00} \\ \hline
	\end{tabular*}
\end{table}
\begin{table}[ht]
	\centering
	\caption{Unbounded collective scheduler without affinity using L=1000,10000 ($640\times640$) 1-byte unsigned 2D tensors and row sorting using quick sort (N=24)}
	\label{tbl:unbounded_collective}
	\footnotesize
	\begin{tabular*}{\columnwidth}{@{\extracolsep{\fill}}|l|l|l|c|c|c|}
		\hline
		Scheduler & L & k & $\sigma(k)$ & $\sigma E(k)$ & $\alpha(k)$ \\ \hline
		collective & 1000     & 1000 & 0.978 & $9.78 \times 10^{-4}$ & 1.00 \\ \hline
		collective & 10000  & 10000 & 0.975 & $9.75 \times 10^{-5}$ & 1.00 \\ \hline
	\end{tabular*}
\end{table}

\begin{figure}[ht] 
	\centering
	
	\begin{minipage}{0.48\textwidth}
		\centering
		\includegraphics[width=\linewidth]{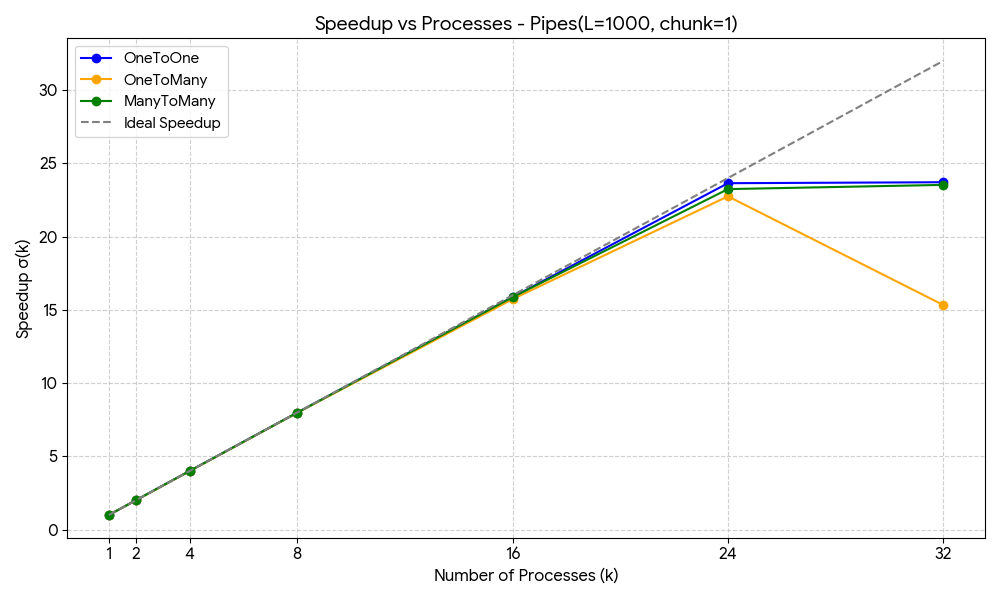} 
		\caption{Performance scaling for all pipe schedulers using L=1000.}
		\label{fig:pipes_L1000}
	\end{minipage}
	\hfill
	\begin{minipage}{0.48\textwidth}
		\centering
		\includegraphics[width=\linewidth]{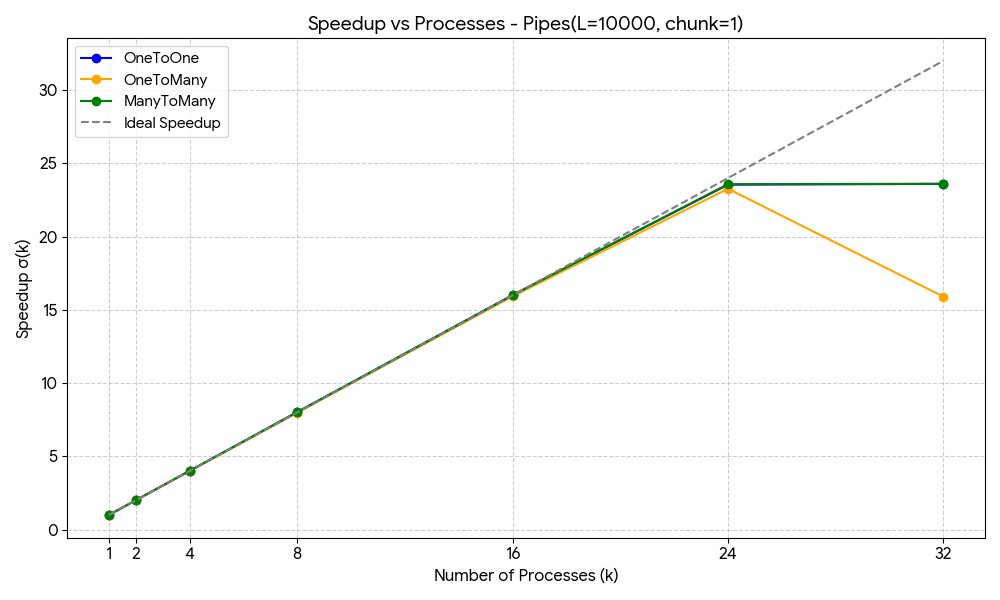}
		\caption{Performance scaling for all pipe schedulers using L=10000.}
		\label{fig:pipes_L10000}
	\end{minipage}
	
\end{figure}

The three pipe-based scheduler designs are presented in Tables~\ref{tbl:pipe11}, \ref{tbl:pipe1m}, and~\ref{tbl:pipemm}, using a chunk size of one, while an indicative scenario examining the optimal chunk size is presented in Tables~\ref{tbl:pipe_cmp_1k} (for $L=1000$) and~\ref{tbl:pipe_cmp_10k} (for $L=10000$) using chunk sizes of $12$ and $105$, respectively. Figures~\ref{fig:pipes_L1000} (for $L=1000$) and~\ref{fig:pipes_L10000} (for $L=10000$) present the speedup results with respect to $k$ for chunk sizes of one. The configuration used was 1-byte unsigned 2D tensors with tasks of row sorting using the quick sort method. In these experiments, the number of processes $k$ is equal to the number of cores $N$, and strict affinity was enforced, ensuring each process is assigned to a specific core.

In pipe experiments, the chunk size is a critical control parameter as it determines the number of frame identifiers transmitted through a pipe per scheduling operation. A chunk size of one provides the finest granularity for load balancing but increases pipe read/write operations, system calls, and scheduling decisions. Conversely, larger chunks reduce communication overhead by transferring more work per pipe transaction; however, they may introduce tail imbalance if one process receives a heavier workload batch while others become idle. Through extensive experimentation with various chunk sizes, we concluded that the optimal chunk size was $12$ for the $L=1000$ scenario and $105$ for the $L=10000$ scenario, as these values maximized performance in terms of execution time (ms). 

Based on the presented results, several key observations emerge regarding the performance characteristics of the three pipe-based scheduler designs. For the one-to-one scheduler (Table~\ref{tbl:pipe11}), we observe near-ideal scaling up to k=16 for both L=1000 and L=10000, with speedup values of 15.875 and 16.012 respectively, corresponding to efficiency values of 0.992 and 1.001. This indicates excellent parallelization efficiency when the number of processes does not exceed the available cores. However, a significant performance degradation occurs when k exceeds the number of physical cores (N=24), with speedup dropping to approximately 23.6 for k=32, resulting in efficiency values of only 0.738-0.741. This degradation is attributed to processor oversubscription, where multiple processes compete for the same core resources, leading to increased context switching and communication overhead.

The one-to-many scheduler (Table~\ref{tbl:pipe1m}) exhibits a distinct performance pattern. For L=1000 with k=24, speedup reaches 22.753 with an efficiency of 0.948, while for L=10000 the corresponding values are 23.270 and 0.970. The most striking observation is the dramatic performance collapse at k=32, where speedup plummets to 15.335 for L=1000 and 15.909 for L=10000, representing efficiencies of only 0.479 and 0.497 respectively. This severe degradation suggests that the one-to-many communication pattern becomes particularly susceptible to resource contention and communication bottlenecks when the system is oversubscribed, likely due to the asymmetric communication patterns where a single producer feeds multiple consumers.

The many-to-many scheduler (Table~\ref{tbl:pipemm}) demonstrates the most robust performance across all configurations. For L=10000, this scheduler achieves nearly perfect scaling up to k=16 with $\sigma(k)$=16.004, and maintains high efficiency at k=24. Even under oversubscription at k=32, the many-to-many scheduler maintains speedup of 23.608 with efficiency 0.738, comparable to the one-to-one design but significantly outperforming the one-to-many scheduler. This superior performance can be attributed to the balanced communication patterns inherent in the many-to-many design, which distributes the communication load more evenly across all participating processes. 

The comparative analysis in Tables~\ref{tbl:pipe_cmp_1k} and~\ref{tbl:pipe_cmp_10k} for k=24 processes with optimized chunk sizes reveals interesting trade-offs. For L=1000 with chunk size 12, all schedulers achieve similar performance, with speedup values ranging from 20.560 to 20.834 and efficiencies between 0.857-0.868, suggesting that larger chunk sizes help amortize communication overhead regardless of the scheduling pattern. For L=10000 with chunk size close to 100, the performance differences become more pronounced: the many-to-many scheduler achieves the highest speedup value of 23.255, closely followed by one-to-one at 23.235 and one-to-many at 23.209. The slightly better performance of the many-to-many scheduler indicates that balanced communication patterns become increasingly beneficial as problem size grows, since a many-to-many communitation allows the producer to balance the tasks in bigger batches (larger chunk sizes) that helps reduce the frequency of communications and provide a more time consuming task-effort per core. This is an indication for a fair task assignment scheduler for heterogenous processing workers.

\begin{table}[ht] 
	\centering
	\setlength{\tabcolsep}{3pt}
		\caption{One-to-one pipe scheduler with affinity using L=1000, 10000 2D tensors (N=24)}
		\label{tbl:pipe11}
		\begin{tabular}{|l|c|c|c|c|c|}
			\hline
			\textbf{Scheduler} & \textbf{$L$} & \textbf{$k$} & \textbf{$\sigma(k)$} & \textbf{$\sigma E(k)$} & \textbf{$\alpha(k)$} \\ \hline
			\textbf{pipe 1:1} & \textbf{1000}  & \textbf{2}  & \textbf{2.005} & \textbf{1.002} & \textbf{-0.002} \\ \hline
			\textbf{pipe 1:1} & \textbf{1000}  & \textbf{4}  & \textbf{4.006} & \textbf{1.002} & \textbf{-0.001} \\ \hline
			pipe 1:1 & 1000 & 8  & 7.976 & 0.997 & 0.000 \\ \hline
			\textbf{pipe 1:1} & \textbf{1000}  & \textbf{16} & \textbf{15.875}& \textbf{0.992} & \textbf{0.001} \\ \hline
			\textbf{pipe 1:1} & \textbf{1000}  & \textbf{24} & \textbf{23.649}& \textbf{0.985} & \textbf{0.001} \\ \hline
			\textbf{pipe 1:1} & \textbf{1000}  & \textbf{32} & \textbf{23.715}& \textbf{0.741} & \textbf{0.011} \\ \hline
			\textbf{pipe 1:1} & \textbf{10000} & \textbf{2}  & \textbf{2.005} & \textbf{1.002} & \textbf{-0.002} \\ \hline
			pipe 1:1 & 10000& 4  & 4.013 & 1.003 & -0.001 \\ \hline
			pipe 1:1 & 10000& 8  & 8.026 & 1.003 & 0.000 \\ \hline
			pipe 1:1 & 10000& 16 & 16.012& 1.001 & 0.000 \\ \hline
			pipe 1:1 & 10000& 24 & 23.547& 0.981 & 0.001 \\ \hline
			pipe 1:1 & 10000& 32 & 23.604& 0.738 & 0.011 \\ \hline
		\end{tabular}
		\vspace{4pt}
		\noindent
		\parbox{\linewidth}{\raggedright \footnotesize \textit{*Note: Values of $\sigma E(k) > 1$ (arising when speedup $\sigma(k) > k$), observed here and in subsequent configurations, indicate super-linear speedup. This is typically caused by the subdivided workloads fitting more efficiently into lower-level CPU caches (L1/L2) during parallel execution.}}
	\hfill
\end{table}
	\begin{table}[ht] 
		
		\centering
		\caption{One-to-many pipe scheduler with affinity using L=1000, 10000 2D tensors (N=24)}
		\label{tbl:pipe1m}
		\begin{tabular}{|l|c|c|c|c|c|}
			\hline
			\textbf{Scheduler} & \textbf{$L$} & \textbf{$k$} & \textbf{$\sigma(k)$} & \textbf{$\sigma E(k)$} & \textbf{$\alpha(k)$} \\ \hline
			pipe 1:M & 1000 & 2  & 1.999 & 1.000 & 0.000 \\ \hline
			pipe 1:M & 1000 & 4  & 3.993 & 0.998 & 0.001 \\ \hline
			pipe 1:M & 1000 & 8  & 7.979 & 0.997 & 0.000 \\ \hline
			pipe 1:M & 1000 & 16 & 15.734& 0.983 & 0.001 \\ \hline
			pipe 1:M & 1000 & 24 & 22.753& 0.948 & 0.002 \\ \hline
			pipe 1:M & 1000 & 32 & 15.335& 0.479 & 0.035 \\ \hline
			pipe 1:M & 10000& 2  & 1.999 & 0.999 & 0.001 \\ \hline
			pipe 1:M & 10000& 4  & 3.994 & 0.999 & 0.001 \\ \hline
			pipe 1:M & 10000& 8  & 7.982 & 0.998 & 0.000 \\ \hline
			pipe 1:M & 10000& 16 & 15.924& 0.995 & 0.000 \\ \hline
			pipe 1:M & 10000& 24 & 23.270& 0.970 & 0.001 \\ \hline
			pipe 1:M & 10000& 32 & 15.909& 0.497 & 0.033 \\ \hline
		\end{tabular}
	\end{table}
	
	
	\begin{table}[ht] 
		\centering
		\caption{Many-to-many pipe scheduler using L=1000, 10000 tensors (N=24)}
		\label{tbl:pipemm}
		\begin{tabular}{|l|c|c|c|c|c|}
			\hline
			\textbf{Scheduler} & \textbf{$L$} & \textbf{$k$} & \textbf{$\sigma(k)$} & \textbf{$\sigma E(k)$} & \textbf{$\alpha(k)$} \\ \hline
			pipe M:M & 1000 & 2  & 2.005 & 1.002 & -0.002 \\ \hline
			pipe M:M & 1000 & 4  & 4.005 & 1.001 & 0.000  \\ \hline
			\textbf{pipe M:M} & \textbf{1000} & \textbf{8}  & \textbf{7.989} & \textbf{0.999} & \textbf{0.000}  \\ \hline
			pipe M:M & 1000 & 16 & 15.868& 0.992 & 0.001  \\ \hline
			pipe M:M & 1000 & 24 & 23.245& 0.969 & 0.001  \\ \hline
			pipe M:M & 1000 & 32 & 23.534& 0.735 & 0.012  \\ \hline
			pipe M:M & 10000& 2  & 2.005 & 1.002 & -0.002 \\ \hline
			\textbf{pipe M:M} & \textbf{10000}& \textbf{4}  & \textbf{4.013} & \textbf{1.003} & \textbf{-0.001} \\ \hline
			pipe M:M & 10000& 8  & 8.026 & 1.003 & 0.000  \\ \hline
			\textbf{pipe M:M}& \textbf{10000}& \textbf{16} & \textbf{16.004}& \textbf{1.000} & \textbf{0.000}  \\ \hline
			\textbf{pipe M:M}& \textbf{10000}& \textbf{24} & \textbf{23.580}& \textbf{0.982} & \textbf{0.001}  \\ \hline
			\textbf{pipe M:M}& \textbf{10000}& \textbf{32} & \textbf{23.608}& \textbf{0.738} & \textbf{0.011}  \\ \hline
		\end{tabular}
	\end{table}
	\begin{table}[ht] 
		\centering
		\caption{Comparative metrics of pipe schedulers for $k = 24$ processes using an $L = 1000$ tensor, with a chunk size of 12 and processor affinity enabled on an $N = 24$ core system.}
		\label{tbl:pipe_cmp_1k}
		\begin{tabular}{|l|c|c|c|c|c|}
			\hline
			\textbf{Scheduler} & \textbf{$L$} & \textbf{$k$} & \textbf{$\sigma(k)$} & \textbf{$\sigma E(k)$} & \textbf{$\alpha(k)$} \\ \hline
			pipe 1:1 & 1000 & 24 & 20.834 & 0.868 & 0.00661 \\ \hline
			pipe M:M & 1000 & 24 & 20.831 & 0.868 & 0.00661 \\ \hline
			pipe 1:M & 1000 & 24 & 20.560 & 0.857 & 0.00725 \\ \hline
		\end{tabular}
	\end{table}
	
	\begin{table}[ht] 
		\caption{Comparative metrics of pipe schedulers for $k = 24$ processes using an $L = 10000$ tensor, with a chunk size of 105 and processor affinity enabled on an $N = 24$ core system.}
		\label{tbl:pipe_cmp_10k}
		\begin{tabular}{|l|c|c|c|c|c|}
			\hline
			\textbf{Scheduler} & \textbf{$L$} & \textbf{$k$} & \textbf{$\sigma(k)$} & \textbf{$\sigma E(k)$} & \textbf{$\alpha(k)$} \\ \hline
			pipe M:M & 10000& 24 & 23.255 & 0.969 & 0.00139 \\ \hline
			pipe 1:1 & 10000& 24 & 23.235 & 0.968 & 0.00144 \\ \hline
			pipe 1:M & 10000& 24 & 23.209 & 0.967 & 0.00148 \\ \hline
		\end{tabular}
	\end{table}


The speedup results presented in Figures~\ref{fig:pipes_L1000} and~\ref{fig:pipes_L10000} for chunk size one visually confirm the trends observed in the tabular data. For both L values, all schedulers show near-linear speedup up to k=16, with the many-to-many scheduler maintaining the best scaling characteristics. The figures clearly illustrate the performance collapse of the one-to-many scheduler at k=32, while the one-to-one and many-to-many schedulers show more graceful degradation. Notably, the performance patterns are remarkably consistent across both L=1000 and L=10000, suggesting that the observed scheduler behavior is primarily determined by the communication patterns and resource allocation strategies rather than the problem size itself. The use of processor affinity ensures that the measured performance reflects the true characteristics of each scheduling design without the confounding effects of process migration.

Based on the comprehensive analysis, the many-to-many (pipe M:M) scheduler emerges as the superior choice for distributed task assignment across all tested configurations. This scheduler consistently demonstrates the most robust performance characteristics, achieving nearly ideal scaling ($\sigma E(k) \ge 0.982$) up to the physical core count (k=24) for both problem sizes, and maintaining the highest efficiency (0.969 for L=10000) even under optimized chunk sizes. The many-to-many design's balanced communication patterns effectively distribute the workload across all participating processes, minimizing bottlenecks and acquiring heterogenous loads. This is particularly evident in the oversubscription scenario (k=32) where the many-to-many scheduler achieves a 48\% improvement over the one-to-many scheduler's performance, demonstrating superior resilience to resource contention.

\subsection{Scenario II. Threads parallelism}\label{s:5_2}
\begin{figure}[ht]
	\centering
	\includegraphics[width=0.4\textwidth, height=6cm, keepaspectratio=false]{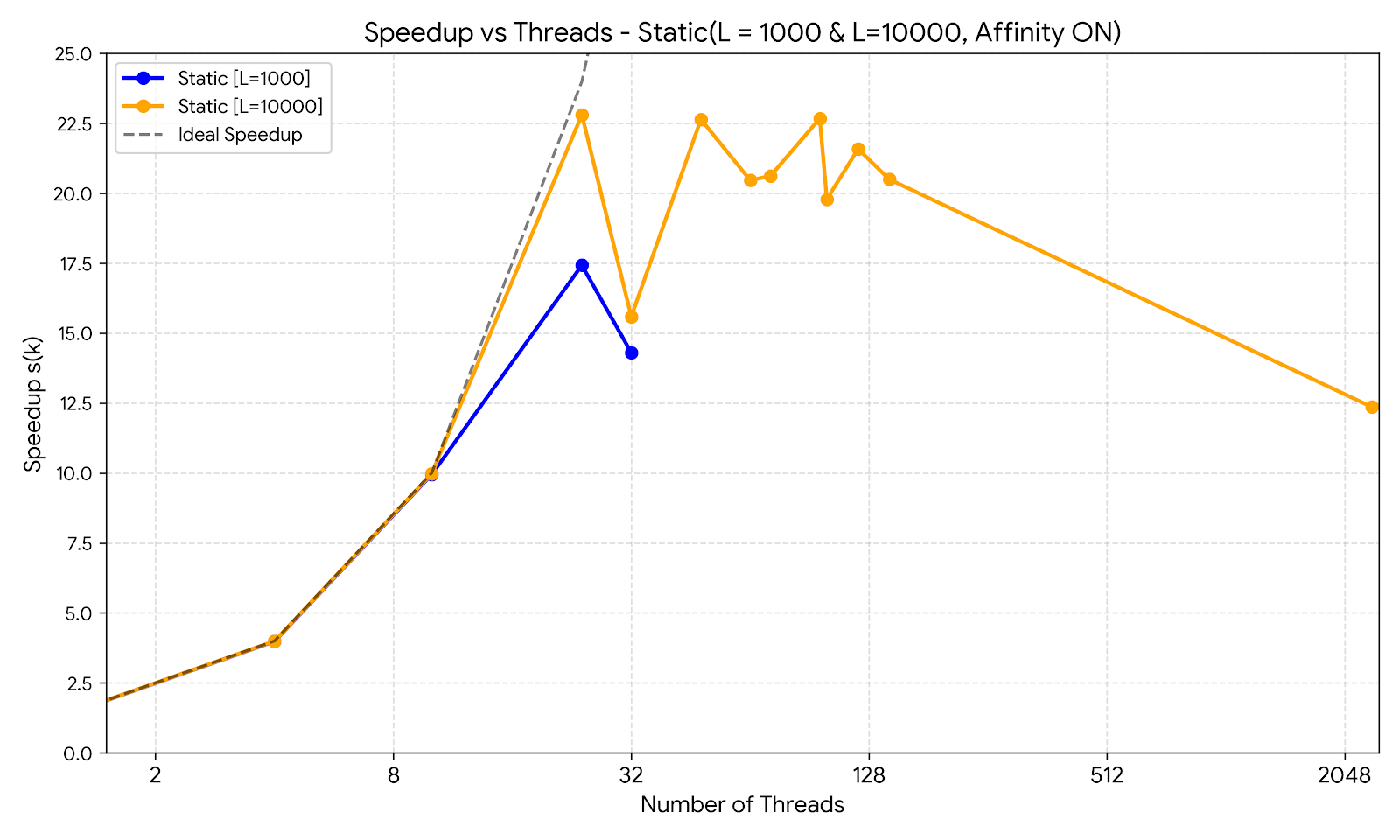}
	\caption{Performance and thread scaling analysis for the Static Scheduler.}
	\label{fig:static_plot}
\end{figure}

\begin{figure}[ht]
	\centering
	\includegraphics[width=0.4\textwidth, height=6cm, keepaspectratio=false]{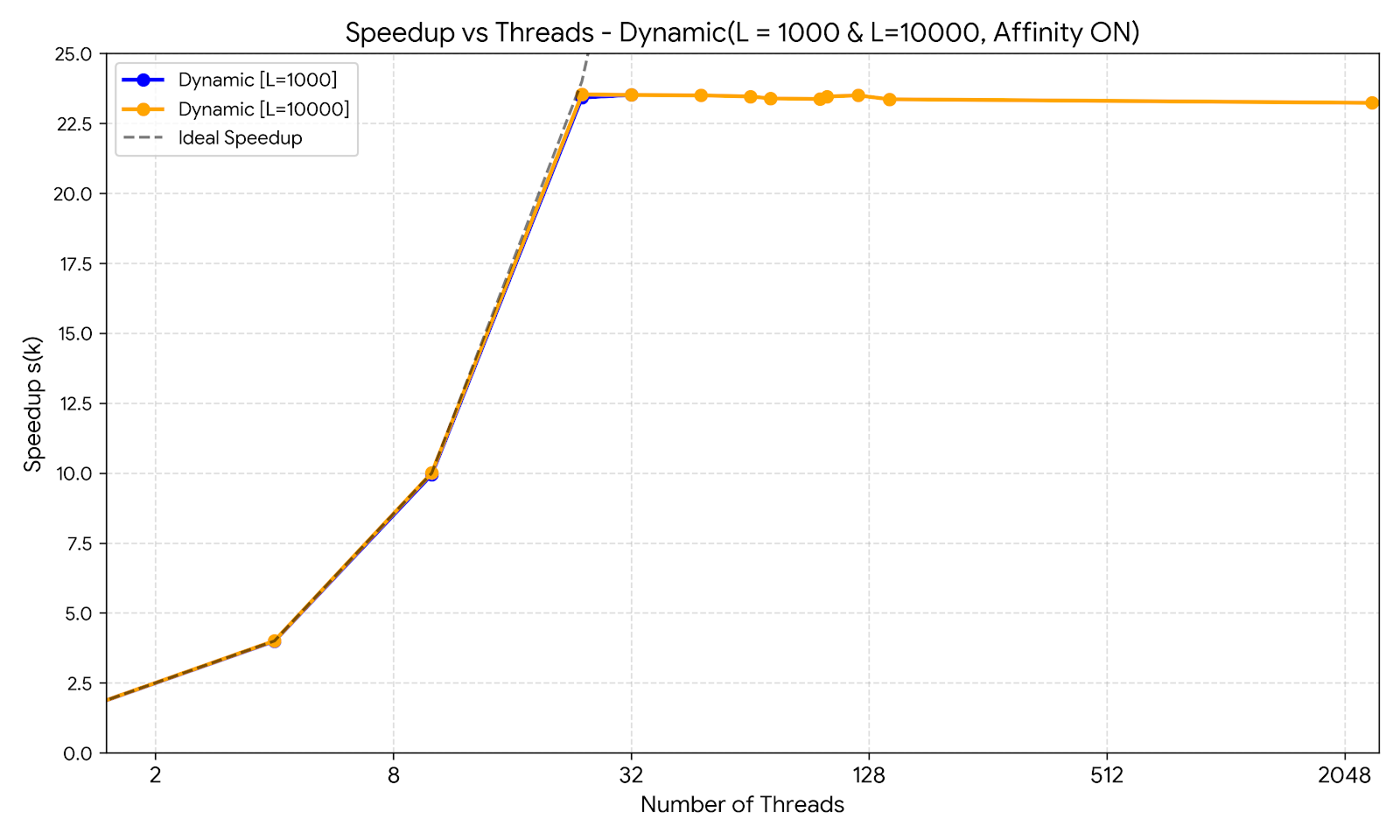}
	\caption{Performance and thread scaling analysis for the Dynamic Scheduler.}
	\label{fig:dynamic_plot}
\end{figure}

\begin{figure}[ht]
\centering
\includegraphics[width=0.4\textwidth, height=6cm, keepaspectratio=false]{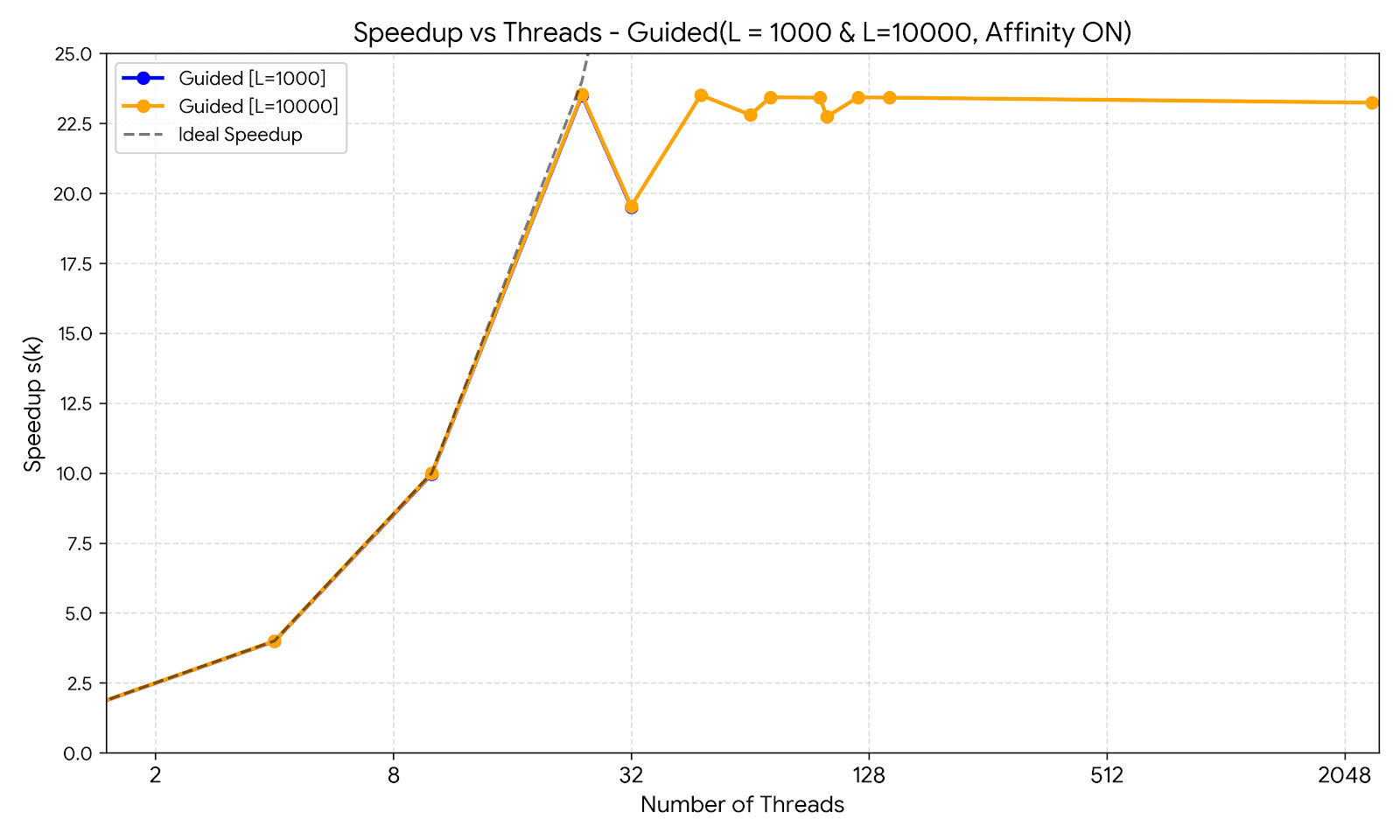}
\caption{Performance and thread scaling analysis for the Guided Scheduler.}
\label{fig:guided_plot}
\end{figure}

\begin{figure}[ht]
	\centering
	\includegraphics[width=0.4\textwidth, height=6cm, keepaspectratio=false]{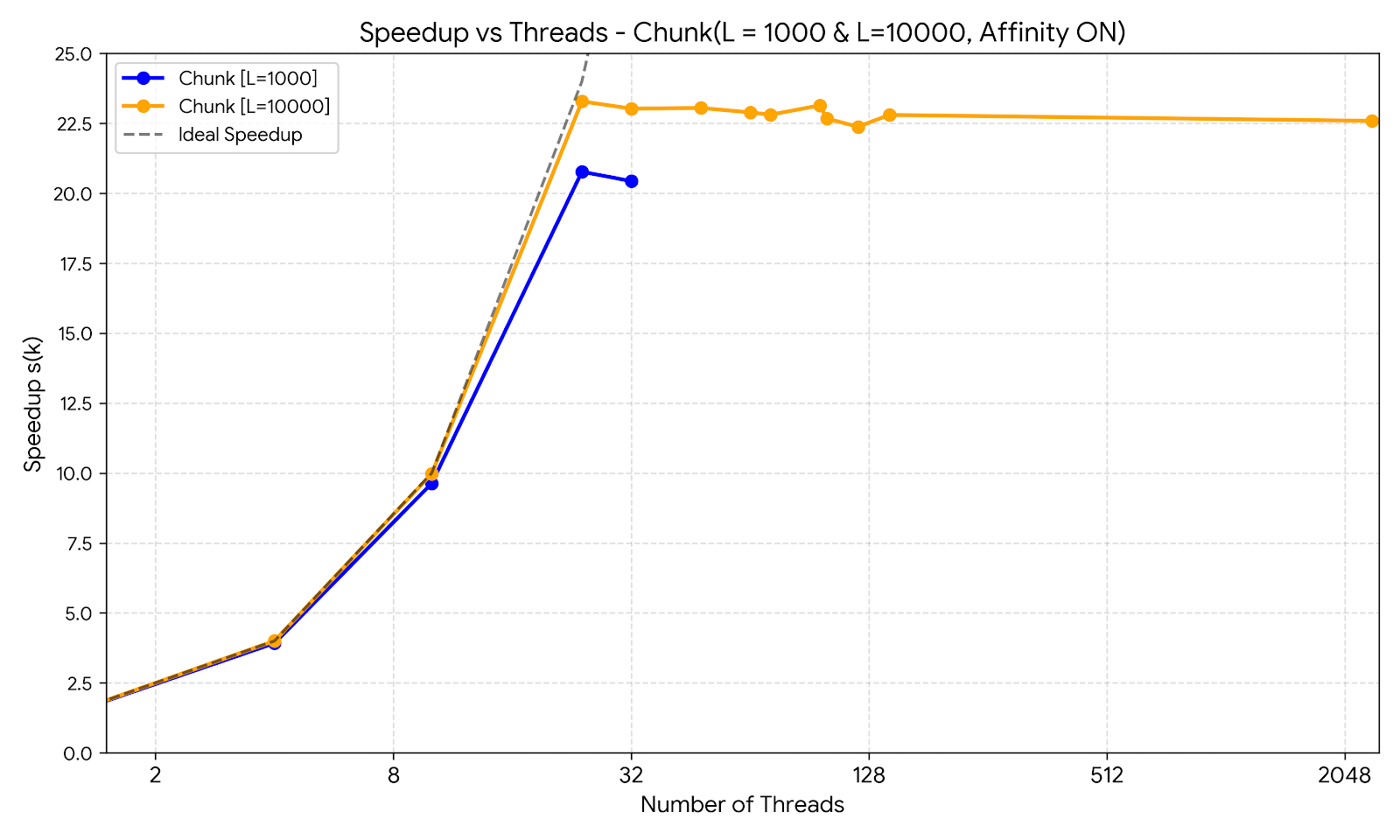}
	\caption{Performance and thread scaling analysis for the Chunk Scheduler.}
	\label{fig:chunk_plot}
\end{figure}

\begin{figure}[ht]
	\centering
	\includegraphics[width=0.4\textwidth, height=6cm, keepaspectratio=false]{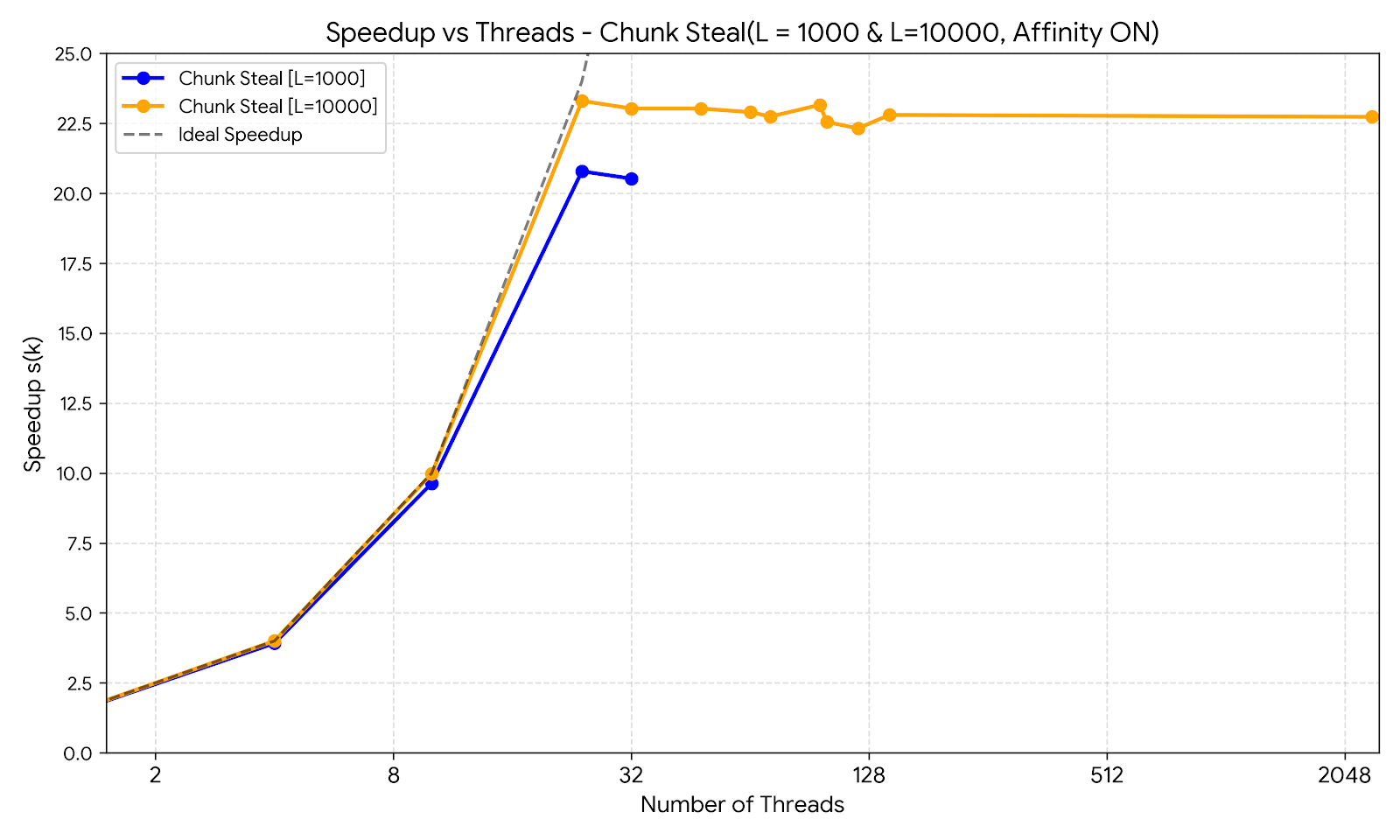}
	\caption{Performance and thread scaling analysis for the Chunk Steal Scheduler.}
	\label{fig:chunk_steal_plot}
\end{figure}

\begin{figure}[ht]
	\centering
	\includegraphics[width=0.4\textwidth, height=6cm, keepaspectratio=false]{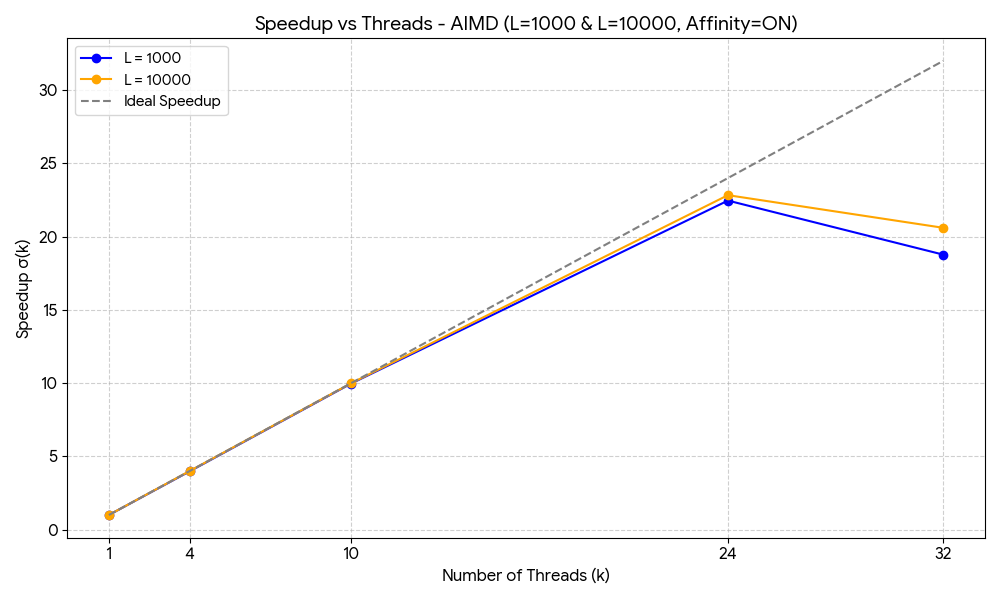}
	\caption{Performance and thread scaling analysis for the AIMD Scheduler.}
	\label{fig:aimd_plot}
\end{figure}

\begin{figure}[ht]
	\centering
	\includegraphics[width=0.4\textwidth, height=6cm, keepaspectratio=false]{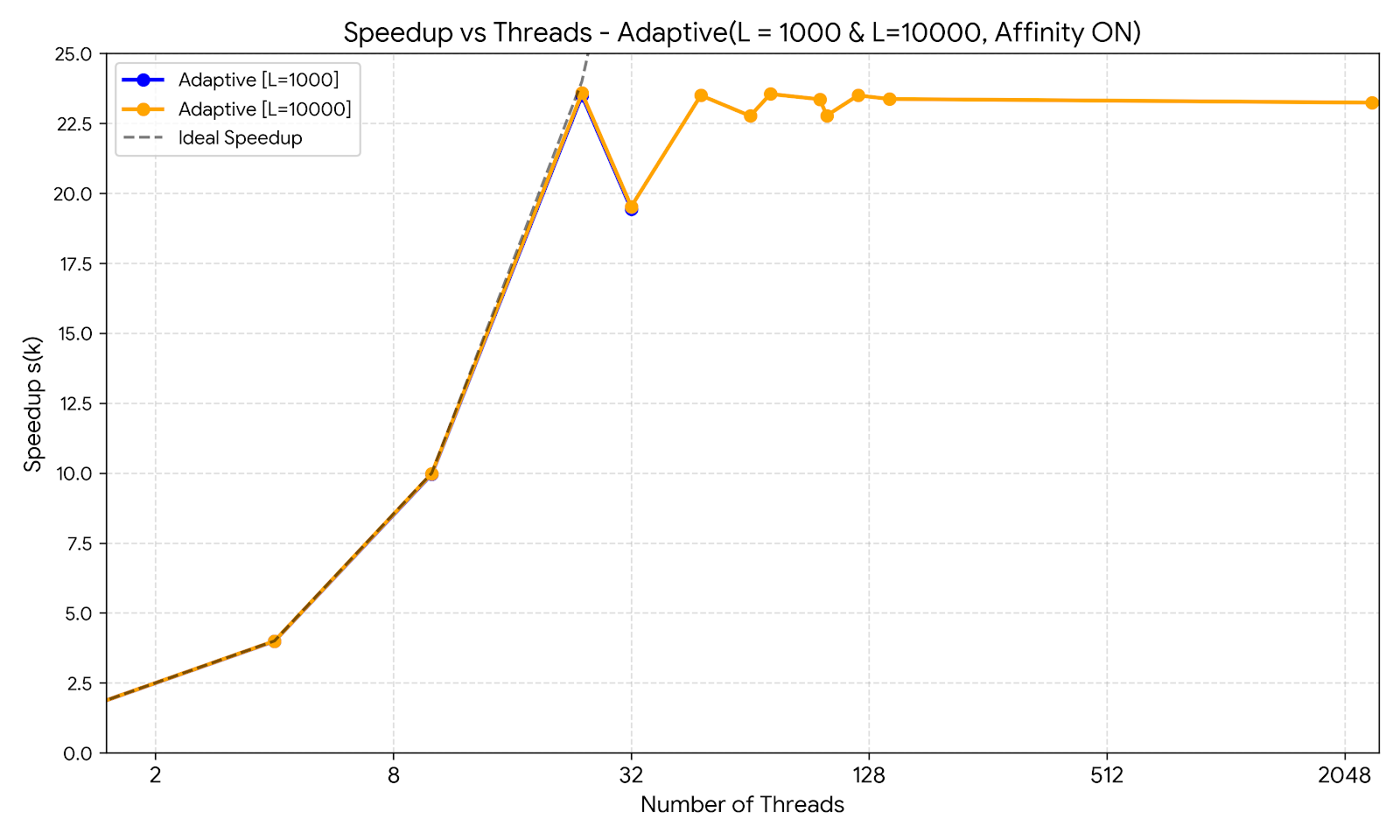}
	\caption{Performance and thread scaling analysis for the Adaptive Chunk Scheduler (ACS).}
	\label{fig:acs_plot}
\end{figure}

Thread scheduling (OpenMP, Pthreads) is tested with $L = 1000, 10000$ tensors ($640 \times 640$) and row sorting ($N = 24$). The static, dynamic, guided, chunk, chunk-stealing, AIMD, and adaptive chunk schedulers are presented in Tables~\ref{tbl:static} through~\ref{tbl:acs} and Figures~\ref{fig:static_plot} through~\ref{fig:acs_plot} with corresponding speedup plots. These figures show how each policy reacts as $k$ increases from low parallelism to the hardware limit and then into oversubscription.
Static scheduling (Table~\ref{tbl:static} and Figure~\ref{fig:static_plot}) assigns work in predetermined portions and therefore has very low scheduling overhead, but faces difficulty if some frames take longer to sort. Dynamic scheduling (Table~\ref{tbl:dynamic} and Figure~\ref{fig:dynamic_plot}) assigns a new task to a thread whenever it finishes its previous one, improving load balance at the cost of more synchronization between workers. Guided scheduling (Table~\ref{tbl:guided} and Figure~\ref{fig:guided_plot}) begins with larger chunks and gradually decreases the chunk size, in order to combine the low overhead of static scheduling with the balancing behavior of dynamic scheduling.
The chunk and chunk-stealing schedulers further test the effect of fixed-size task grouping in the thread family. In the chunk scheduler (Table~\ref{tbl:chunk} and Figure~\ref{fig:chunk_plot}), frames are assigned in fixed batches, using a chunk size of 10 for $L=1000$ and a chunk size of 100 for $L=10000$. Smaller chunks improve balancing but increase mutex lock/scheduler activity, while larger chunks reduce synchronization and thus, risk leaving idle threads near the end. Chunk stealing (Table~\ref{tbl:chunk_steal} and Figure~\ref{fig:chunk_steal_plot}) starts from chunk-based assignment but allows idle threads to take remaining work from other workers, improving performance under imbalance, while adding queue management and stealing overhead.
The AIMD (Table~\ref{tbl:aimd} and Figure~\ref{fig:aimd_plot}) and adaptive chunk (Table~\ref{tbl:acs} and Figure~\ref{fig:acs_plot}) schedulers extend this idea by changing the amount of active or assigned work according to runtime feedback, but the figures show how this extra control logic doesn't improve speedup compared to the simpler dynamic and guided approaches in our implementation.

\begin{table}[ht]
	\centering
	\caption{Static scheduler with affinity using L=1000,10000 ($640\times640$) 1-byte unsigned 2D tensors and row sorting using quick sort (N=24) }
	\label{tbl:static}
	\begin{tabular}{|p{1.3cm}|m{1.7cm}|m{0.7cm}|m{0.75cm}|m{0.7cm}|m{1cm}|}
		\hline
		Scheduler & L & k & $\sigma(k)$ & $\sigma E(k)$ & $\alpha(k)$ \\ \hline
		static & 1000   & 4 & 3.985 & 0.996 & 0.00133 \\ \hline
		static & 1000     & 10 & 9.946 & 0.994 & 0.00067 \\ \hline
		static & 1000     & 24 & 17.426 & 0.726 & 0.0164 \\ \hline
		static & 1000     & 32 & 14.3 & 0.446 & 0.04007 \\ \hline
		static & 10000   & 4 & 3.987 & 0.999 & 0.00033 \\ \hline
		static & 10000  & 10 & 9.949 & 0.994 & 0.00067 \\ \hline
		static & 10000  & 24 & 22.958 & 0.956 & 0.002 \\ \hline
		static & 10000  & 32 & 15.599 & 0.487 & 0.03398 \\ \hline
	\end{tabular}
\end{table}

\begin{table}[ht]
	\centering
	\caption{Dynamic scheduler with affinity using L=1000,10000 ($640\times640$) 1-byte unsigned 2D tensors and row sorting using quick sort (N=24)}
	\label{tbl:dynamic}
	\begin{tabular}{|p{1.3cm}|m{1.7cm}|m{0.7cm}|m{0.75cm}|m{0.7cm}|m{1cm}|}
		\hline
		Scheduler & L & k  & $\sigma(k)$ & $\sigma E(k)$ & $\alpha(k)$ \\ \hline
		\textbf{dynamic} & \textbf{1000}   & \textbf{4} & \textbf{3.992} & \textbf{0.998} & \textbf{0.00067} \\ \hline
		dynamic & 1000   & 10 & 9.945 & 0.994 & 0.00067 \\ \hline
		dynamic & 1000     & 24 & 23.421 & 0.975 & 0.00111 \\ \hline
		\textbf{dynamic} & \textbf{1000}     & \textbf{32} & \textbf{23.518} & \textbf{0.734} & \textbf{0.01169} \\ \hline
		\textbf{dynamic} & \textbf{10000}   & \textbf{4} & \textbf{4.001} & \textbf{1.0002} & \textbf{0.00007} \\ \hline
		\textbf{dynamic} & \textbf{10000}   & \textbf{10} & \textbf{10.005} & \textbf{1.0005} & \textbf{0.00005} \\ \hline
		\textbf{dynamic} & \textbf{10000}  & \textbf{24} & \textbf{23.706} & \textbf{0.9870} & \textbf{0.00057} \\ \hline
		\textbf{dynamic} & \textbf{10000}  & \textbf{32} & \textbf{23.358} & \textbf{0.729} & \textbf{0.01199} \\ \hline
	\end{tabular}
\end{table}

\begin{table}[H]
	\centering
	\caption{Guided scheduler with affinity using L=1000,10000 ($640\times640$) 1-byte unsigned 2D tensors and row sorting using quick sort (N=24) }
	\label{tbl:guided}
	\begin{tabular}{|p{1.3cm}|m{1.7cm}|m{0.7cm}|m{0.75cm}|m{0.7cm}|m{1cm}|}
		\hline
		Scheduler & L & k & $\sigma(k)$ & $\sigma E(k)$ & $\alpha(k)$ \\ \hline
		guided & 1000   & 4 & 3.988 & 0.997 & 0.001 \\ \hline
		guided & 1000   & 10 & 9.965 & 0.996 & 0.00039 \\ \hline
		\textbf{guided} & \textbf{1000}     & \textbf{24} & \textbf{23.497} & \textbf{0.979} & \textbf{0.00093} \\ \hline
		guided & 1000     & 32 & 19.507 & 0.61 & 0.02066 \\ \hline
		guided & 10000  & 4 & 3.983 & 0.995 & 0.00055 \\ \hline
		guided & 10000   & 10 & 9.961 & 0.996 & 0.00044 \\ \hline
		guided & 10000  & 24 & 23.662 & 0.985 & 0.00066 \\ \hline
		guided & 10000  & 32 & 19.479 & 0.608 & 0.0208 \\ \hline
	\end{tabular}
\end{table}

\begin{table}[ht]
	\centering
	\caption{Chunk(10/100) scheduler using L=1000,10000 ($640\times640$) 1-byte unsigned 2D tensors and row sorting (N=24)}
	\label{tbl:chunk}
	\begin{tabular}{|p{1.5cm}|m{1.7cm}|m{0.5cm}|m{0.75cm}|m{0.75cm}|m{1cm}|}
		\hline
		Scheduler & L & k  & $\sigma(k)$ & $\sigma E(k)$ & $\alpha(k)$ \\ \hline
		chunk(10) & 1000   & 4 & 3.918 & 0.979 & 0.00701 \\ \hline
		chunk(10) & 1000     & 10 & 9.632 & 0.963 & 0.00424 \\ \hline
		chunk(10) & 1000     & 24 & 20.77 & 0.865 & 0.00676 \\ \hline
		chunk(10) & 1000     & 32 & 20.436 & 0.639 & 0.01825 \\ \hline
		\textbf{chunk(100)} & \textbf{10000}   & \textbf{4} & \textbf{4.001} & \textbf{1.0002} & \textbf{0.00007} \\ \hline
		chunk(100) & 10000  & 10 & 9.954 & 0.995 & 0.00055 \\ \hline
		chunk(100) & 10000  & 24 & 23.443 & 0.976 & 0.00106 \\ \hline
		chunk(100) & 10000  & 32 & 23.353 & 0.729 & 0.01199 \\ \hline
	\end{tabular}
\end{table}

\begin{table}[ht]
	\centering
	\caption{Chunk steal(10/100) scheduler using L=1000,10000 ($640\times640$) 1-byte unsigned 2D tensors and row sorting (N=24)}
	\label{tbl:chunk_steal}
	\begin{tabular}{|p{1.7cm}|m{1.7cm}|m{0.5cm}|m{0.75cm}|m{0.75cm}|m{1cm}|}
		\hline
		Scheduler & L & k  & $\sigma(k)$ & $\sigma E(k)$ & $\alpha(k)$ \\ \hline
		chunk\_s(10) & 1000   & 4 & 3.917 & 0.979 & 0.00707 \\ \hline
		chunk\_s(10) & 1000     & 10 & 9.634 & 0.963 &  0.00423\\ \hline
		chunk\_s(10) & 1000     & 24 & 20.787 & 0.866 & 0.00672 \\ \hline
		chunk\_s(10) & 1000     & 32 & 20.524 & 0.641 & 0.01804 \\ \hline
		\textbf{chunk\_s(100)} & \textbf{10000}   & \textbf{4} & \textbf{4.004} & \textbf{1.001} & \textbf{0.00033} \\ \hline
		chunk\_s(100) & 10000  & 10 & 9.954 & 0.995 & 0.00055 \\ \hline
		chunk\_s(100) & 10000  & 24 & 23.376 & 0.974 & 0.00116 \\ \hline
		chunk\_s(100) & 10000  & 32 & 23.005 & 0.718 & 0.01266 \\ \hline
	\end{tabular}
\end{table}

\begin{table}[ht]
	\centering
	\caption{AIMD scheduler , using L=1000,10000 ($640\times640$) 1-byte unsigned 2D tensors and row sorting (N=24) }
	\label{tbl:aimd}
	\begin{tabular}{|p{1.3cm}|m{1.7cm}|m{0.7cm}|m{0.75cm}|m{0.75cm}|m{1.2cm}|}
		\hline
		Scheduler & L & k  & $\sigma(k)$ & $\sigma E(k)$ & $\alpha(k)$ \\ \hline
		AIMD & 1000     & 4 & 3.979 & 0.994 & 0.002 \\ \hline
		\textbf{AIMD} & \textbf{1000}     & \textbf{10} & \textbf{9.972} & \textbf{0.997} & \textbf{0.00033} \\ \hline
		AIMD & 1000     & 24 & 22.462 & 0.935 & 0.00302 \\ \hline
		AIMD & 1000     & 32 & 18.772 & 0.586 & 0.02278 \\ \hline
		\textbf{AIMD} & \textbf{10000}  & \textbf{4} & \textbf{4} & \textbf{1} & \textbf{0} \\ \hline
		AIMD & 10000  & 10 & 9.993 & 0.999 & 0.00224 \\ \hline
		AIMD & 10000  & 24 & 22.832 & 0.951 & 0.00011 \\ \hline
		AIMD & 10000  & 32 & 20.604 & 0.643 & 0.01791 \\ \hline
	\end{tabular}
\end{table}

\begin{table}[ht]
	\centering
	\caption{Adaptive chunk scheduler using L=1000,10000 ($640\times640$) 1-byte unsigned 2D tensors and row sorting (N=24)}
	\label{tbl:acs}
	\begin{tabular}{|p{1.3cm}|m{1.7cm}|m{0.7cm}|m{0.75cm}|m{0.7cm}|m{1cm}|}
		\hline
		Scheduler & L & k & $\sigma(k)$ & $\sigma E(k)$ & $\alpha(k)$ \\ \hline
		ACS & 1000   & 4 & 3.99 & 0.998 & 0.00079 \\ \hline
		ACS & 1000   & 10 & 9.957 & 0.996 & 0.00047 \\ \hline
		ACS & 1000     & 24 & 23.495 & 0.979 & 0.00093 \\ \hline
		ACS & 1000     & 32 & 19.442 & 0.608 & 0.02084 \\ \hline
		\textbf{ACS} & \textbf{10000}   & \textbf{4} & \textbf{4.004} & \textbf{1.001} & \textbf{0.00033} \\ \hline
		ACS & 10000  & 10 & 9.946 & 0.994 & 0.00067 \\ \hline
		ACS & 10000  & 24 & 23.565 & 0.981 & 0.00084 \\ \hline
		ACS & 10000  & 32 & 19.453 & 0.607 & 0.02088 \\ \hline
	\end{tabular}
\end{table}

\section{Experimental Results and Discussion}\label{s:6}
\subsection{Process-Based Scheduler Results}\label{s:6_1}

Among the process-based schedulers evaluated in Scenario
5.1, the bounded prolific and bounded collective schedulers
both exhibited strong scalability near the hardware parallelism limit. The bounded prolific scheduler achieved the
highest process-family result, reaching speedup efficiency
close to 0.96 for the larger workload at k = 24. The bounded
collective scheduler followed closely, confirming that both
fork-based process strategies can scale well when the workload is sufficiently large and the number of workers remains
close to the available logical processors.

The bounded prolific scheduler slightly outperformed
the collective approach in the reported measurements because its parent-centered process management and cyclic
task assignment introduce less coordination overhead.
More specifically, for L = 1000, the prolific scheduler was
approximately 1.5\% better than the collective scheduler,
while for L = 10000, it was approximately 2.4\% better.
Overall, the prolific scheduler was approximately 2\% better than the collective scheduler. The collective scheduler
distributes process creation more broadly, which can reduce
parent-side spawning pressure, but it also requires additional process tracking and synchronization. For the evaluated row-sorting workload, this extra coordination cost
slightly reduced its advantage.

The three pipe-based schedulers provide a different
process-execution model. Instead of assigning all work statically at process creation time, they stream task identifiers
to worker processes through inter-process communication
channels. This makes the pipe schedulers more dynamic
than the bounded prolific and bounded collective schedulers, but it also introduces pipe read/write overhead and
more scheduling decisions during execution. The one-to-one pipe scheduler uses a single shared communication pipe
from the producer to all workers. Its advantage is simplicity: the parent writes work identifiers into one channel, and
workers consume tasks from that shared stream. However,
because all workers compete for the same pipe, contention
may appear when the number of workers increases. This
design is therefore simple and stable, but less flexible under
high concurrency.

The one-to-many pipe scheduler uses one communication pipe per worker. This removes much of the worker-side
contention on a single shared pipe because the parent can
send work directly to a specific process. Its performance
depends strongly on how well the parent distributes chunks
among workers. If the distribution is balanced, dedicated
pipes reduce communication conflicts. If the distribution is
imperfect, however, some workers may finish earlier while
others still hold larger remaining chunks.

The relative ranking among pipe schedulers depends on the workload size. For L=1000, the one-to-one pipe scheduler gives the best result among the pipe family, marginally outperforming the many-to-many pipe scheduler at k=24, with a difference of approximately 1.7\%. For L=10000, the many-to-many pipe scheduler achieves the best result, marginally outperforming the one-to-one pipe scheduler at k=24,with a difference of approximately 0.1\%. Overall, performance differences within the pipe family are small, and the choice between the one-to-one and many-to-many designs depends primarily on the workload size.

The many-to-many pipe scheduler's advantage in larger workloads (L=10000) comes from using separate communication paths for task assignment and completion notification. Apart from receiving tasks , workers also report when they finish. This feedback gives the parent process a more accurate runtime view of worker availability, allowing it to push new tasks toward idle workers more quickly. As workload size increases, the benefits of this feedback loop outweigh its added IPC complexity. For smaller workloads (L=1000), however, the simpler one-to-one pipe scheduler design avoids the extra coordination overhead and achieves slightly better results.

Chunk size is a central factor in the performance of all
three pipe-based schedulers. In this workload, each task
corresponds to one tensor frame, and a chunk is a group of
consecutive frame identifiers sent through the pipe as one
scheduling unit. With a small chunk size, the scheduler has
fine-grained control: workers receive smaller batches, finish
them quickly, and can be reassigned new work with minimal waiting. This improves load balance, especially when
some frames take longer to sort than others. The cost is that
the system performs more pipe writes, pipe reads, context
switches, and scheduling decisions.

With a larger chunk size, each pipe transaction transfers more useful work. This lowers IPC overhead because
the parent communicates less frequently with the workers.
Larger chunks can therefore improve performance when
the workload is regular and each chunk has similar cost.
However, if the chunk becomes too large, load imbalance
appears: one worker may receive a heavy batch and continue executing while other workers become idle. Thus, the
chunk parameter creates a direct trade off between communication overhead and load-balancing accuracy. The best
chunk size is not simply the smallest or largest value; it is
the value that keeps pipe traffic low without allowing long
idle periods.

Overall, the process-based results show that when the
same task size is used per worker, specifically one frame per
worker, bounded prolific scheduling is the strongest of the
classic fork-based approaches. However, for the pipe implementations, the most important performance factor turned
out to be the discovery of the best task size. In the L = 1000
scenario, the best task size was 12 frames per worker, while
in the L = 10000 scenario, it was 105 frames per worker.
The prolific scheduler wins through simplicity and low coordination overhead, while the many-to-many pipe scheduler wins inside the pipe family because its completion feedback mechanism gives it better runtime adaptability.

\subsection{Thread-Based Scheduler Results}\label{s:6_2}
Among the thread-based schedulers evaluated in Scenario
5.2, the dynamic scheduler achieved the strongest overall
performance. For L = 1000, the guided scheduler was approximately 0.3\% better than the dynamic scheduler. For L
= 10000, the dynamic scheduler was approximately 1.4\%
better than the chunk-steal scheduler. 

Dynamic scheduling performs well because threads repeatedly request new work as soon as they finish their previous task. This keeps the worker pool active and limits
idle time, especially when the execution cost of individual
frames varies. Its main cost is synchronization on the shared
scheduling state, but for this workload the overhead remains
small enough that the scheduler maintains near-ideal scaling
up to the hardware parallelism limit.

Guided scheduling also performs strongly because it
starts with larger chunks and gradually reduces chunk size
as the remaining work decreases. This reduces scheduling overhead during the early part of execution, while still
improving load balance near the end. In practice, guided
scheduling behaves like a compromise between static and
dynamic scheduling: it avoids the high scheduling frequency of fully dynamic work distribution while still preventing the severe tail imbalance that can appear in purely
static assignment.

The static scheduler performs well when the workload is
large and regular, but it is less robust when work distribution becomes uneven. Since each thread receives a predetermined portion of the workload, any imbalance in task cost
can leave some threads idle while others continue working.
This explains why static scheduling can be competitive in
some cases but does not match the overall robustness of dynamic scheduling.

The chunk scheduler reduces scheduling overhead by
assigning groups of frames at a time. Its behavior depends heavily on the chosen chunk size. Small chunks
improve balance but increase scheduling overhead, while
large chunks reduce synchronization frequency but risk idle
workers near the end of execution. This mirrors the chunk size trade off observed in the pipe schedulers, although the
cost here is thread synchronization rather than pipe-based
IPC.

The chunk-stealing scheduler attempts to correct imbalance by allowing idle workers to steal work from others.
While this improves theoretical load balance, it also introduces queue-management and stealing overhead. For the
evaluated row-sorting workload, the extra synchronization
cost reduces its practical advantage, although it remains
highly competitive for the larger L = 10000 workload.

The guided scheduler and the adaptive chunk scheduler (ACS) produce very similar results throughout the experiments. At $k=24$, guided reaches $\sigma=23.497$ for $L=1000$ and $\sigma=23.662$ for $L=10000$, while ACS reaches $\sigma=23.495$ and $\sigma=23.565$ respectively. These differences are well within the 1\% noise threshold. The reason guided remains competitive is that the evaluated workload --- row-wise quicksort on frames of identical dimensions --- is sufficiently homogeneous in task cost that the per-thread speed-factor adaptation performed by ACS provides little practical benefit. The source of irregularity is the data-dependent behavior of quicksort (different frames contain differently ordered data and therefore take slightly different times to sort), but this variation is small and not correlated across consecutive frames; as a result, the guided scheduler's decreasing-chunk strategy already captures most of the available load-balancing gain without the additional measurement and adaptation overhead of ACS.

The AIMD scheduler is introduced as an exploratory, contention-aware policy designed to ensure fairness and prevent system overload over long-running executions in multi-tenant platforms. It regulates concurrency through a contention window controlled by processor-utilization feedback. However, the results show it is generally weaker than the dynamic and guided schedulers for the current benchmark. This underperformance can be attributed to two main factors. First, the experimental scenario executes in an isolated environment without unpredictable external background contention; therefore, the AIMD throttling mechanism occasionally restricts active worker threads when the system could otherwise safely sustain full concurrency, effectively wasting available CPU cycles. Second, to enforce this contention window, the scheduler relies on heavier synchronization mechanisms---specifically condition variables---to continuously block and wake threads based on the number of in-flight chunks. This introduces significant synchronization and wake-up overhead compared to the simple, fast mutex locks used by the dynamic scheduler. Consequently, while AIMD provides a valuable safety mechanism for shared systems, its conservative throttling and synchronization overheads make it suboptimal for dedicated, high-performance batch processing.

\subsection{Affinity versus Non-Affinity Execution}\label{s:6_3}
Processor affinity did not consistently produce significant performance improvements across all evaluated configurations. For several workloads, affinity-enabled and non-affinity executions exhibited comparable execution times, suggesting that the Linux scheduler already performs effective runtime load balancing and processor migration for the evaluated workloads. This behavior is consistent with the design of modern work-stealing and process-based schedulers, which are capable of approximating near-optimal core utilization without explicit affinity constraints under low-to-moderate contention.

Nevertheless, affinity remained beneficial in scenarios where execution reproducibility and reduced variability were prioritized over raw throughput. In particular, affinity-enabled configurations demonstrated more consistent worker-to-core placement, which is especially relevant for higher worker counts and process-based schedulers operating under deterministic timing requirements. In oversubscribed configurations, affinity also served to limit unnecessary processor migrations and cache invalidations, thereby preserving cache locality and reducing the associated performance degradation.

\subsection{Scalability and Serialization Analysis}\label{s:6_4}
In this work, scalability is defined as the ability of a scheduler to maintain increasing speedup as the contention level approaches the affinity-supported limit. The measured speedup and efficiency values indicate that several schedulers scaled efficiently up to the hardware parallelism limit of N = 24 processing units, with performance gains remaining consistent across increasing worker counts within this range.

As worker counts exceeded the available hardware parallelism, oversubscription effects increasingly introduced synchronization pressure, context-switching overhead, and cache-locality degradation. These factors collectively resulted in reduced parallel efficiency and increased estimated serialization fractions $\alpha$, reflecting the rising proportion of execution time attributed to non-parallelizable overhead, including lock contention and inter-core communication latency.

Schedulers with lower synchronization and communication overhead generally maintained better scalability under increasing concurrency levels, particularly for the regular frame-level row-sorting workload evaluated in this study.

\subsection{Results Summary}\label{s:6_5}
Combining the measurements from Section 5 with the analysis of Sections 6.1 and 6.2, the best schedulers per family and their relative performance are summarized below. 

\begin{itemize} 
	\item \textbf{Best fork-based process scheduler:} Bounded prolific. 
	\item \textbf{Best pipe process scheduler:} 
	\begin{itemize} 
		\item For $L=1000$: one-to-one pipe scheduler (small number of low computational tasks). 
		\item For $L=10000$: many-to-many pipe scheduler (big number of low and medium computational tasks). 
	\end{itemize} 
	\item \textbf{Best overall process scheduler} (fork and pipe combined): 
	\begin{itemize} 
		\item For $L=1000$, $k=24$: one-to-one pipe ($\sigma=23.649$), exceeding bounded prolific (small number of low computational tasks) ($\sigma=22.394$) and also marginally exceeding the best thread scheduler (dynamic, $\sigma=23.421$). 
		\item For $L=10000$, $k=24$: many-to-many pipe ($\sigma=23.580$), exceeding bounded prolific scheduler. ($\sigma=22.940$) and also marginally exceeding dynamic ($\sigma=23.706$ -- dynamic scheduler leads at $L=10000$). 
	\end{itemize} 
	\item \textbf{Best thread scheduler:} dynamic (particularly strong under oversubscription) with guided as a near-equal alternative at $k=24$. 
\end{itemize} 

\paragraph{Process-thread schedulers comparison:} 
\begin{itemize} 
	\item $k=24$, $L=1000$: one-to-one pipe ($\sigma=23.649$) vs.\ dynamic ($\sigma=23.421$); pipe leads by $\approx 1.0\%$. 
	\item $k=24$, $L=10000$: dynamic ($\sigma=23.706$) vs.\ many-to-many pipe ($\sigma=23.580$); thread leads by $\approx 0.5\%$. 
	\item $k=32$, $L=1000$: dynamic ($\sigma=23.518$) vs.\ one-to-one pipe ($\sigma=23.715$); pipe leads by $\approx 0.8\%$. 
	\item $k=32$, $L=10000$: dynamic ($\sigma=23.358$) vs.\ many-to-many pipe ($\sigma=23.608$); pipe leads by $\approx 1.1\%$. 
\end{itemize} 

These differences are all within or close to the 1\% noise threshold, confirming that at hardware parallelism ($k=24$) and at modest oversubscription ($k=32$), the best pipe schedulers and the best thread schedulers perform essentially on par. The clear differentiation appears only for fork-based process schedulers (bounded prolific and collective), which are substantially outperformed under oversubscription. 

\paragraph{Process-Pipe schedulers comparison:} 
\begin{itemize} 
	\item $k=24$, $L=1000$: one-to-one pipe is $\approx 5.6\%$ better than bounded prolific. 
	\item $k=24$, $L=10000$: many-to-many pipe is $\approx 2.8\%$ better than bounded prolific. 
	\item $k=32$, $L=1000$: one-to-one pipe is $\approx 53\%$ better than bounded prolific. 
	\item $k=32$, $L=10000$: many-to-many pipe is $\approx 48.5\%$ better than bounded prolific. 
\end{itemize} 

Across all evaluated configurations, dynamic thread scheduling and the best pipe schedulers match each other closely at hardware parallelism ($k=24$), and both substantially outperform fork-based process schedulers under oversubscription ($k=32$), confirming that lightweight execution units coordinated through dynamic IPC are preferable for the evaluated shared-memory row-sorting workload. 

Table \ref{tab:runtimesummary} presents a consolidated view of the absolute execution times ($T_1$ and $T_{24}$) alongside the achieved speedup for the best-performing process and thread-based schedulers at hardware parallelism ($k=24$). While relative metrics like speedup and efficiency highlight the theoretical scalability of each approach, raw execution times provide necessary context regarding the absolute computational scale and the practical impact of system overheads (e.g., process forking, pipe communication, and mutex synchronization). As observed in the absolute parallel runtimes ($T_{24}$), the one-to-one pipe scheduler achieves the best overall performance for the smaller workload ($L=1000$), finishing marginally faster than the dynamic thread scheduler. Conversely, for the larger workload ($L=10000$), the dynamic thread scheduler achieves the lowest absolute execution time, demonstrating superior efficiency and lower overhead as the workload size increases.

\begin{table}[H]
	\centering
	\caption{Sequential ($T_{1}$) and parallel ($T_{24}$) execution times together with the corresponding speedup of the best-performing process- and thread-based schedulers.}
	\label{tab:runtimesummary}
	
	\begin{tabular}{|p{1.5cm}|p{1.2cm}|m{0.8cm}|m{0.8cm}|m{0.8cm}|m{0.8cm}|}
		\hline
		\textbf{Scheduler} &
		\textbf{Category} &
		\textbf{Load} &
		\textbf{$T_{1}$ (ms)} &
		\textbf{$T_{24}$ (ms)} &
		\parbox{0.7cm}{\textbf{\scriptsize Speedup ($\sigma$)}} \\ \hline
		
		One-to-one Pipe  & Process & 1000  & 37564 & 1588 & 23.65 \\ \hline
		Dynamic          & Thread  & 1000  & 37652 & 1608 & 23.42 \\ \hline
		Many-to-many Pipe& Process & 10000 & 37580 & 1617 & 23.25 \\ \hline
		Dynamic          & Thread  & 10000 & 37657 & 1589 & 23.71 \\ \hline
		
	\end{tabular}
\end{table}

\subsection{Limitations of the Experimental Workload}\label{s:6_6}
While the evaluated schedulers demonstrate clear performance trends, it is important to acknowledge that these conclusions are drawn from a specific and relatively narrow workload. The experimental scenario relies entirely on row-wise quick-sort operations applied to uniform three-dimensional tensors (where every frame has identical $640 \times 640$ dimensions). 

Because the computational geometry of the assigned tasks is perfectly uniform, the source of workload irregularity is not structural (i.e., varying task sizes), but rather algorithmic. Specifically, the execution time of the quick-sort algorithm is highly data-dependent, varying with the initial permutation and randomness of the values within each row. This introduces fine-grained, unpredictable, stochastic timing variations across frames. 

The observed advantage of the dynamic and guided thread schedulers must be interpreted within this context. These schedulers excel here because they effectively smooth out this specific type of data-driven, micro-irregularity without incurring high overhead. For workloads with severe structural irregularities (e.g., sparse matrix processing or processing frames of dynamically varying resolutions), or conversely, for workloads with strictly predictable and constant computational costs (e.g., simple vector addition), the relative performance and ranking of these schedulers might differ. Consequently, broadening these conclusions to entirely different computational patterns remains a subject for future investigation.

\section{Conclusion}\label{s:7}
This paper compared several process-based, pipe-based, and thread-based schedulers for a shared-memory row-sorting workload on an x86-64 many-core platform using large three-dimensional unsigned 8-bit tensors stored in System V shared memory. 

The central finding is that dynamic and guided thread scheduling and selected pipe schedulers (one-to-one for $L=1000$, many-to-many for $L=10000$) perform very close to the ideal speedup up to $k=24$. At hardware parallelism, the best pipe and best thread schedulers are essentially on par, with differences below 1\%. 
Oversubscription ($k=32$) penalizes fork-based process schedulers (bounded prolific and collective) far more strongly than either thread or pipe schedulers: fork-based speedup collapses to roughly $15$--$16$ at $k=32$, while dynamic threads and one-to-one/many-to-many pipes sustain speedup above $23$. 

Among fork-based process schedulers, bounded prolific slightly outperforms bounded collective due to simpler process management and lower coordination overhead. 
Pipe schedulers remain competitive and can match or marginally exceed thread schedulers at $k=24$, particularly with tuned chunk sizes; the many-to-many design benefits from completion feedback at larger workload sizes. 
Thread-based schedulers are preferable overall because they avoid the heavier costs of process creation and inter-process communication; process and pipe schedulers remain valuable when isolation, explicit communication, or distributed-style execution is required. 

Future work includes extending the evaluation to distributed MPI systems, NUMA-aware scheduling strategies, heterogeneous CPU/GPU environments, larger or more irregular workloads, and chunk-based pipe schedulers using chunk sizes greater than one task per communication operation.
\label{sec:con}

{
    \small
	\bibliographystyle{IEEEtranS}
    \bibliography{main}

@article{agathos22,
	title = {Compiler-assisted, adaptive runtime system for the support of {OpenMP} in embedded multicores},
	volume = {110},
	issn = {0167-8191},
	url = {https://www.sciencedirect.com/science/article/pii/S0167819122000035},
	doi = {10.1016/j.parco.2022.102895},
	journal = {Parallel Computing},
	author = {Agathos, Spiros N. and Dimakopoulos, Vassilios V. and Kasmeridis, Ilias K.},
	month = {May},
	year = {2022},
	pages = {102895},
}

@inproceedings{clet-ortega14,
	address = {Cham},
	title = {Evaluation of {OpenMP} {Task} {Scheduling} {Algorithms} for {Large} {NUMA} {Architectures}},
	isbn = {978-3-319-09873-9},
	doi = {10.1007/978-3-319-09873-9_50},
	booktitle = {Euro-{Par} 2014 {Parallel} {Processing}},
	publisher = {Springer International Publishing},
	author = {Clet-Ortega, Jérôme and Carribault, Patrick and Pérache, Marc},
	editor = {Silva, Fernando and Dutra, Inês and Santos Costa, Vítor},
	year = {2014},
	pages = {596--607}
}

@article{booth22,
  author  = {Joshua Dennis Booth and Phillip Allen Lane},
  title   = {An adaptive self-scheduling loop scheduler},
  journal = {Concurrency and Computation: Practice and Experience},
  volume  = {34},
  number  = {6},
  pages   = {e6750},
  year    = {2022},
  doi     = {10.1002/cpe.6750}
}

@inproceedings{guo10,
  author    = {Yi Guo and Jisheng Zhao and Vincent Cave and Vivek Sarkar},
  title     = {SLAW: A Scalable Locality-aware Adaptive Work-stealing Scheduler},
  booktitle = {Proceedings of the 24th IEEE International Parallel and Distributed Processing Symposium (IPDPS)},
  year      = {2010},
  pages     = {1--12},
  doi       = {10.1109/IPDPS.2010.5470425}
}

@article{akturk2019,
	title = {Adaptive {Thread} {Scheduling} in {Chip} {Multiprocessors}},
	volume = {47},
	issn = {0885-7458, 1573-7640},
	url = {http://link.springer.com/10.1007/s10766-019-00637-y},
	doi = {10.1007/s10766-019-00637-y},
	number = {5-6},
	journal = {International Journal of Parallel Programming},
	author = {Akturk, Ismail and Ozturk, Ozcan},
	month = {Dec},
	year = {2019},
	pages = {1014--1044}
	}

@incollection{thoman12,
  author    = {Peter Thoman and Karl Raffenetti and Thomas Fahringer},
  title     = {Automatic OpenMP Loop Scheduling: A Combined Compiler and Runtime Approach},
  booktitle = {OpenMP in a Heterogeneous World},
  series    = {Lecture Notes in Computer Science},
  publisher = {Springer},
  year      = {2012},
  doi       = {10.1007/978-3-642-30961-8_7}
}

@article{muller25,
  author={Muller Korndorfer, Jonas H. and Mohammed, Ali and Eleliemy, Ahmed and Guilloteau, Quentin and Krummenacher, Reto and Ciorba, Florina M.},
  journal={IEEE Access}, 
  title={A Comparative Study of OpenMP Scheduling Algorithm Selection Strategies}, 
  year={2025},
  volume={13},
  number={},
  pages={151216-151234},
  doi={10.1109/ACCESS.2025.3602234}}

@article{KarpFlatt1990,
  author  = {Alan H. Karp and Horace P. Flatt},
  title   = {Measuring Parallel Processor Performance},
  journal = {Communications of the ACM},
  volume  = {33},
  number  = {5},
  pages   = {539--543},
  year    = {1990},
  doi     = {10.1145/78607.78614}
}

@inproceedings{Amdahl1967,
  author    = {Gene M. Amdahl},
  title     = {Validity of the Single Processor Approach to Achieving Large-Scale Computing Capabilities},
  booktitle = {Proceedings of the April 18--20, 1967, Spring Joint Computer Conference},
  series    = {AFIPS '67 (Spring)},
  year      = {1967},
  pages     = {483--485},
  publisher = {Association for Computing Machinery},
  doi       = {10.1145/1465482.1465560}
}

@article{Lawrynowicz11,
  author  = {Anna Lawrynowicz},
  title   = {Advanced scheduling with genetic algorithms in supply networks},
  journal = {Journal of Manufacturing Technology Management},
  year    = {2011},
  volume  = {22},
  number  = {6},
  pages   = {748--769},
  doi     = {10.1108/17410381111149620}
}

@misc{vasileiadou25,
	address = {Rochester, NY},
	type = {{SSRN} {Scholarly} {Paper}},
	title = {Investigating {High}-{Performance} {Computing} of {Matrix} {Multiplications} using {Processes}, {Threads}, {OpenMP}, and {MPI}: {A} {Case} {Study} of {Experiments} on {X86}, {ARM}, and {ARM} {Clusters}},
	shorttitle = {Investigating {High}-{Performance} {Computing} of {Matrix} {Multiplications} using {Processes}, {Threads}, {OpenMP}, and {MPI}},
	url = {https://papers.ssrn.com/abstract=5350041},
	doi = {10.2139/ssrn.5350041},
	publisher = {Elsevier SSRN},
	author = {Vasileiadou, Alexandra and Gkenoudi, Virginia and Kiose, Nikoleta and Konstantinopoulou, Antonia and Michou, Glykeria and Psaropa, Maria and Spanoudaki, Konstantina and Tsoumani, Meropi and Kontogiannis, Sotirios},
	month = {Jul},
	year = {2025}
}

@inproceedings{Sysel20,
  author    = {Martin Sysel},
  title     = {A Comparison of Processes and Threads Creation},
  booktitle = {Software Engineering Perspectives in Intelligent Systems},
  editor    = {Radek Silhavy and Petr Silhavy and Zuzana Prokopova},
  series    = {Advances in Intelligent Systems and Computing},
  volume    = {1294},
  pages     = {990--997},
  publisher = {Springer, Cham},
  year      = {2020},
  doi       = {10.1007/978-3-030-63322-6_85}
}

@inproceedings{jeffay93,
  author    = {Kevin Jeffay},
  title     = {The Real-Time Producer/Consumer Paradigm: A Paradigm for the Construction of Efficient, Predictable Real-Time Systems},
  booktitle = {Proceedings of the 1993 ACM Symposium on Applied Computing},
  pages     = {796--804},
  year      = {1993},
  publisher = {ACM},
  doi       = {10.1145/162754.168703}
}

@incollection{dijkstra68,
  author    = {Edsger W. Dijkstra},
  title     = {Co-operating Sequential Processes},
  booktitle = {Programming Languages: NATO Advanced Study Institute, Lectures Given at a Three Weeks Summer School Held in Villard-le-Lans, 1966},
  editor    = {F. Genuys},
  pages     = {43--112},
  publisher = {Academic Press},
  address   = {London},
  year      = {1968},
  isbn      = {0-12-279750-7},
  url       = {https://pure.tue.nl/ws/files/4279816/344354178746665.pdf}
}

@article{kontogiannis11,
  title   = {A probing algorithm with Adaptive Workload Load Balancing capabilities for heterogeneous clusters},
  volume  = {3},
  journal = {Journal of Computing},
  author  = {Kontogiannis, S. and Karakos, Alexandros},
  month   = {Jan},
  year    = {2011},
  pages   = {107--116},
  url     = {https://www.researchgate.net/publication/3841626_A_probing_algorithm_with_Adaptive_Workload_Load_Balancing_capabilities_for_heterogeneous_clusters}
}

@inproceedings{Mohtasham15,
  author    = {Amin Mohtasham and Joao Barreto},
  title     = {Fair Adaptive Parallelism for Concurrent Transactional Memory Applications},
  booktitle = {Proceedings of the 27th ACM Symposium on Parallelism in Algorithms and Architectures},
  series    = {SPAA '15},
  pages     = {68--70},
  year      = {2015},
  publisher = {Association for Computing Machinery},
  doi       = {10.1145/2755573.2755609}
}

@article{Gustafson88,
  author  = {John L. Gustafson},
  title   = {Reevaluating Amdahl's Law},
  journal = {Communications of the ACM},
  volume  = {31},
  number  = {5},
  pages   = {532--533},
  year    = {1988},
  doi     = {10.1145/42411.42415}
}

@misc{OpenMP52Spec,
  title        = {OpenMP Application Programming Interface Version 5.2},
  author       = {{OpenMP Architecture Review Board}},
  organization = {OpenMP Architecture Review Board},
  year         = {2021},
  month        = {Nov},
  url          = {https://www.openmp.org/wp-content/uploads/OpenMP-API-Specification-5-2.pdf}
}

@misc{gunther_general_2008,
	title = {A {General} {Theory} of {Computational} {Scalability} {Based} on {Rational} {Functions}},
	url = {http://arxiv.org/abs/0808.1431},
	doi = {10.48550/arXiv.0808.1431},
	urldate = {2026-04-16},
	publisher = {arXiv},
	author = {Gunther, Neil J.},
	month = {Aug},
	year = {2008},
	note = {arXiv:0808.1431 [cs]}
	}

@inproceedings{gunther1993usl,
  author    = {Neil J. Gunther},
  title     = {A Simple Capacity Model of Massively Parallel Transaction Systems},
  booktitle = {Proceedings of the 19th International Computer Measurement Group Conference},
  pages     = {1035--1044},
  year      = {1993},
  address   = {San Diego, CA, USA},
  url       = {https://www.perfdynamics.com/Papers/njgCMG93.pdf}
}

@manual{mpi41standard,
  title        = {MPI: A Message-Passing Interface Standard, Version 4.1},
  author       = {{Message Passing Interface Forum}},
  organization = {MPI Forum},
  year         = {2023},
  month        = {Nov},
  url          = {https://www.mpi-forum.org/docs/mpi-4.1/mpi41-report.pdf}
}

@book{gropp1999usingmpi,
  author    = {William Gropp and Ewing Lusk and Anthony Skjellum},
  title     = {Using MPI: Portable Parallel Programming with the Message-Passing Interface},
  edition   = {2},
  publisher = {MIT Press},
  address   = {Cambridge, MA},
  year      = {1999},
  isbn      = {9780262571326}
}

@article{blumofe1999,
  author  = {Robert D. Blumofe and Charles E. Leiserson},
  title   = {Scheduling multithreaded computations by work stealing},
  journal = {Journal of the ACM},
  volume  = {46},
  number  = {5},
  pages   = {720--748},
  month   = sep,
  year    = {1999},
  doi     = {10.1145/324133.324234},
  url     = {https://doi.org/10.1145/324133.324234}
}

@book{bovet2005,
  author    = {Daniel P. Bovet and Marco Cesati},
  title     = {Understanding the {Linux} Kernel},
  edition   = {3},
  publisher = {O'Reilly Media},
  address   = {Sebastopol, CA},
  year      = {2005}
}

@article{hoare1978,
  author  = {C. A. R. Hoare},
  title   = {Communicating sequential processes},
  journal = {Communications of the ACM},
  volume  = {21},
  number  = {8},
  pages   = {666--677},
  month   = aug,
  year    = {1978},
  doi     = {10.1145/359576.359585},
  url     = {https://doi.org/10.1145/359576.359585}
}

@manual{posix2017,
  title        = {The Open Group Base Specifications Issue 7, 2018 edition, {IEEE} Std 1003.1-2017},
  author       = {{IEEE} and {The Open Group}},
  year         = {2018},
  url          = {https://pubs.opengroup.org/onlinepubs/9699919799/}
}

@inproceedings{jacobson1988,
   author    = {Van Jacobson},
  title     = {Congestion Avoidance and Control},
  booktitle = {Proceedings of ACM SIGCOMM},
  year      = {1988},
  pages     = {314--329},
  doi       = {10.1145/52324.52356},
  url       = {https://dl.acm.org/doi/10.1145/52324.52356}
}

@article{chiu1989,
    author  = {Dah-Ming Chiu and Raj Jain},
  title   = {Analysis of the Increase/Decrease Algorithms for Congestion Avoidance in Computer Networks},
  journal = {Computer Networks and ISDN Systems},
  volume  = {17},
  number  = {1},
  pages   = {1--14},
  year    = {1989},
  url     = {https://doi.org/10.1016/0169-7552(89)90019-6}
}

@inproceedings{kontogiannis05,
    author    = {Kontogiannis, S. and Mamatas, L. and Psaras, I. and Tsaoussidis, V.},
  title     = {Measuring Transport Protocol Potential for Energy Efficiency},
  booktitle = {Wired/Wireless Internet Communications},
  publisher = {Springer},
  address   = {Berlin, Heidelberg},
  pages     = {333--342},
  year      = {2005},
  doi       = {10.1007/11424505_32},
  isbn      = {978-3-540-32104-0},
  url       = {https://doi.org/10.1007/11424505_32}
}

@article{kontogiannis2011probing,
  author  = {Kontogiannis, Sotirios and Karakos, Alexandros S.},
  title   = {A Probing Algorithm with {Adaptive Workload Load Balancing} Capabilities for Heterogeneous Clusters},
  journal = {Journal of Computing},
  volume  = {3},
  number  = {7},
  pages   = {107--116},
  month   = jul,
  year    = {2011},
  url     = {https://www.researchgate.net/profile/S-Kontogiannis/publication/236648853_A_probing_algorithm_with_Adaptive_Workload_Load_Balancing_capabilities_for_heterogeneous_clusters/links/66027134a4857c796282ddf5/A-probing-algorithm-with-Adaptive-Workload-Load-Balancing-capabilities-for-heterogeneous-clusters.pdf}
}

@article{kontogiannis2014albl,
  author  = {Kontogiannis, Sotirios and Karakos, Alexandros},
  title   = {{ALBL}: An Adaptive Load Balancing Algorithm for Distributed Web Systems},
  journal = {International Journal of Communication Networks and Distributed Systems},
  volume  = {13},
  number  = {2},
  pages   = {144--168},
  year    = {2014},
  doi     = {10.1504/IJCNDS.2014.064041},
  url     = {https://doi.org/10.1504/IJCNDS.2014.064041}
}
}
\end{document}